\documentclass[12pt, a4paper]{article}
\pdfoutput=1
\usepackage{amsmath}
\usepackage{amsfonts}
\usepackage{amssymb}
\usepackage{graphicx, rotating}
\usepackage{epsfig}
\usepackage{latexsym}
\usepackage{graphicx}
\usepackage{color}
\usepackage{amsmath,bm,amssymb}
\usepackage{cite}
\usepackage{slashed}
\usepackage{hyperref}
\hypersetup{colorlinks, citecolor=bluscuro, linkcolor=black, urlcolor=bluscuro}
\definecolor{rossos}{cmyk}{0,1,1,0.55}
\definecolor{bluscuro}{rgb}{0.15, 0.2, .85}
\definecolor{bluchiaro}{cmyk}{1,.3,0.,0.1}

\numberwithin{equation}{section}

\newcommand{\be}{\begin{equation}}
\newcommand{\ee}{\end{equation}}
\newcommand{\bea}{\begin{eqnarray}}
\newcommand{\eea}{\end{eqnarray}}

\def\simlt{\stackrel{<}{{}_\sim}}

\def\pp{{\scriptscriptstyle +}}
\def\mm{{\scriptscriptstyle -}}

\newcommand{\arXiv}[2]{\href{http://arxiv.org/pdf/#1}{{\tt [#2/#1]}}}
\newcommand{\arXivold}[1]{\href{http://arxiv.org/pdf/hep-#1}{{\tt [#1]}}}

\def\bma#1{\mbox{\boldmath{$#1$}}}

\begin{document}
\allowdisplaybreaks
%FRONTPAGE2%%%%%%
\begin{titlepage}
\begin{flushright}
IFT-UAM/CSIC-21-69
\end{flushright}
\vspace{.3in}

\vspace{1cm}
\begin{center}
{\Large\bf\color{black} 
Pseudo-Bounces vs. New Instantons}
\\
\vspace{1cm}{
{\large J.R.~Espinosa} and {\large J. Huertas}
%\vspace{0.3cm}
} \\[7mm]
{\it  Instituto de F\'{\i}sica Te\'orica UAM/CSIC, \\ 
C/ Nicol\'as Cabrera 13-15, Campus de Cantoblanco, 28049, Madrid, Spain}

\end{center}
\bigskip

\vspace{.4cm}

\begin{abstract}
Some false vacua do not decay via bounces. This usually happens when a flat direction of the tunneling action due to scale invariance is lifted to a sloping valley by a scale breaking perturbation, pushing the bounce off to infinity. We compare two types of alternative decay configurations that have been proposed recently to describe decay in such cases: pseudo-bounces and new instantons. Although both field configurations are quite similar, we find that the pseudo-bounce action is lower than the new instanton one and describes more faithfully the bottom of the action valley. In addition, pseudo-bounces cover a range of field space wider than new instantons
and, as a result, lead to an action that can be lower than the one for new instantons by orders of magnitude.
\end{abstract}
\bigskip

\end{titlepage}

\section{Divergent Rates and Decays Without Bounce\label{sec:intro}}

False vacua are a common occurrence in particle physics models
and quite relevant for cosmology. Their decay proceeds by the nucleation by quantum tunneling of a bubble of the more stable phase that expands and transforms space to the deeper vacuum phase. For long-lived vacua, the decay rate per unit volume is exponentially suppressed as
\be
\Gamma/V = A\  e^{-S/\hbar}\ ,
\label{rate}
\ee
where $A$ has dimensions of $[$energy$]^4$ and is set by the typical mass scale of the potential while the key quantity is the tunneling action $S$. The general expression (\ref{rate}) holds provided $S/\hbar\gg 1$, in which case the semiclassical approximation applies.

The calculation of $S$ in quantum field theory can be conveniently
done using a Euclidean approach, pioneered by Coleman\cite{Coleman}, and ultimately justified by its agreement with the WKB approach. It goes as follows. Consider a  real scalar field  $\varphi$ with potential $V(\varphi)$ having a false vacuum at $\varphi_+$ and a true minimum at $\varphi_-$. In the absence of gravity, set $V(\varphi_+)=0$. The tunneling action $S$ for the decay of the  $\varphi_+$ vacuum is calculated by taking imaginary time and so working in 4d Euclidean space. One finds an $O(4)$-symmetric field configuration $\varphi_B(\rho)$ (or Euclidean bounce) that connects $\varphi_+$ (at Euclidean time $\tau\to-\infty$) with a bubble-like configuration that probes the basin of the true vacuum at $\varphi_-$ (at $\tau=0$) and goes back to $\varphi_+$ (at $\tau\to\infty$). This bounce is a saddle point of  the Euclidean action for $\varphi$:
\be
S_E[\varphi] = 2\pi^2\int_0^\infty  \left[\frac12 \dot{\varphi}^2 + V(\varphi)\right]\rho^3d\rho\ .
\label{SE}
\ee
and so it solves the Euler-Lagrange (or bounce) equation:
\be
\ddot{\varphi} +\frac{3}{\rho}\dot{\varphi} = V'\ ,
\label{EoM4}
\ee
where a dot (prime) stands for a derivative with respect to $\rho$ ($\varphi$). The boundary conditions are
\be
\dot\varphi_B(0)=0\ ,\quad \varphi_B(\infty)=\varphi_+\ .
\ee
Taking $\rho$ as time and $\varphi$ as position, Eq.~(\ref{EoM4}) can be interpreted as governing the motion of a particle in the inverted potential $-V(\varphi)$ with a velocity-dependent and time-decreasing friction force. The bounce solution can be searched for by scanning the value of $\varphi_B(\rho=0)$, until the boundary condition at $\rho\rightarrow \infty$ is satisfied. The tunneling action for the decay of the false vacuum $\varphi_+$ is precisely $S=S_E[\varphi_B]$.

When $V(\varphi)$ has a true vacuum there is generically a bounce solution (but not always, see below), that can be found by the  undershoot/overshoot method \cite{Coleman}.  
In trying for solutions of (\ref{EoM4}), if $\varphi(0)$ has $-V(\varphi(0))<0$, friction implies that $\varphi(\rho)$ does not reach $\varphi_+$: an undershot. If instead one starts with $\varphi(0)$ very close to the true minimum $\varphi_-$, the field  spends a long time rolling very slowly away from $\varphi_-$, friction dies off and (Euclidean) energy conservation ensures that  $\varphi(\rho)$ reaches $\varphi_+$ with non zero velocity: an overshot. By continuity, there is a $\varphi(0)$ leading to $\varphi(\infty)=\varphi_+$ (thus corresponding to the bounce), that can therefore be found by interval bisection.

However, there are potentials that escape the previous simple expectation. As is well known, in a negative quartic potential, $V=-\lambda\varphi^4/4$, the false vacuum at $\varphi=0$ can decay not via a single bounce but via an infinite family of Fubini bounces \cite{Fubini,Lipatov} $\varphi(\rho)=\varphi_e/(1+\rho^2/R^2)$ with arbitrary $\varphi_e$ [and radius $R^2=8/(\lambda\varphi_e^2)$] all having the same action $S=8\pi^2/(3\lambda)$. 
The indeterminacy of $\varphi_e$ and the constancy of $S$ both follow from the scale invariance of the potential.\footnote{
Scale invariance implies that, if $\varphi_B(\rho)$ is a bounce solution,
then the rescaled field configuration $a\,\varphi_B(ar)$ (with $a>0$) is also a bounce solution. As seen explicitly from the form of the Fubini bounce, the $a$-rescaling amounts to a rescaling of $\varphi_e\rightarrow a\varphi_e$ (or to a rescaling of the radius $R\rightarrow R/a$).}
That is, the action functional $S_E[\varphi]$ has a flat direction
in field configuration space along the family of Fubini bounces.

Still, there are other potentials (with a false vacuum) for which there is no Euclidean bounce.  As a trivial example take a piece-wise
potential with a false vacuum at some $\varphi_+<0$, with a barrier that reaches its maximum at $\varphi=0$, while for $\varphi>0$,
$V(\varphi)=-\lambda\varphi^4/4$ \cite{NoBounce}. Any solution of equation (\ref{EoM4}) with $\varphi(0)>0$ and $\dot\varphi(0)=0$ ends at $\varphi(\infty)=0$, without reaching $\varphi_+<0$, so no bounce exists.

There is a powerful method to prove the absence of the bounce
\cite{Affleck,ESM}. Let us illustrate it with the potential
\be
V(\varphi)=\frac12 m^2\varphi^2-\frac{\lambda}{4}\varphi^4+\frac{1}{\Lambda^2}\varphi^6\ ,
\label{VmlL}
\ee
with $\Lambda^2\gg m^2>0$. For $m^2/\Lambda^2<\lambda^2/32$,  there is a metastable vacuum at $\varphi_\pp=0$ separated by a barrier from the true vacuum at $\varphi_\mm$ with $\varphi_\mm^2=\Lambda^2\left[\lambda+\sqrt{\lambda^2-24m^2/\Lambda^2}\right]/12$. Let $\varphi_B(\rho)$ be the bounce configuration for the decay out of $\varphi_+$. Consider the rescaled field profile $\varphi_a(\rho)\equiv a \varphi_B(a\rho)$, with Euclidean action (after rescaling the integration variable)
\be
S_E[\varphi_a]=2\pi^2\int_0^\infty \left[\frac12 \left(\frac{d\varphi_B}{d\rho}\right)^2-\frac14 \lambda\varphi_B^4\right]\rho^3d\rho +2\pi^2
\int_0^\infty \left(\frac{1}{2a^2} m^2\varphi_B^2+\frac{a^2}{\Lambda^2}\varphi_B^6\right)\rho^3d\rho \ .
\ee
As $\varphi_B(\rho)$ extremizes the Euclidean action, one should have $dS_E[\varphi_a]/da=0$ at $a=1$, which gives
\be
\int_0^\infty \left(-\frac12 m^2\varphi_B^2+\frac{1}{\Lambda^2}\varphi_B^6\right)\rho^3d\rho=0\ .
\label{rescaling}
\ee 
Fulfilling this integral condition requires a cancellation between quartic and sextic terms. If either is absent, the condition implies $\varphi_B(\rho)\equiv 0$ and there is no bounce.\footnote{In other words, if $m^2\neq 0$ but $\Lambda\rightarrow\infty$ there are only undershots. If $m^2=0$ with a finite $\Lambda$, there are only overshots. In this latter case there is a true minimum, but no bounce.}
The general form of the integral constraint that a bounce should satisfy was presented in \cite{ESM} and reads
\be
\int_0^\infty \sum_\alpha\frac{\partial V(\varphi_B)}{\partial p_\alpha}d_{(\alpha)} p_\alpha  \rho^3d\rho=0\ ,
\label{BCond}
\ee
where $ p_\alpha$ are the parameters in the potential $V$, with mass dimensions $d_{(\alpha)}$. 

Even when there is no bounce, the false vacuum decays: quantum fluctuations of the field in the false vacuum continue to nucleate bubbles that probe the deep part of the potential with decay rates that depend on the shape of the bubble. But how should one calculate the overall decay rate?
This problem was recently revisited in \cite{PseudoB} where pseudo-bounces were introduced to describe how false vacua that cannot decay via proper bounces could still decay. In the Euclidean formalism, pseudo-bounces
are Euclidean field configurations that have a homogeneous core, with the field sitting at a field value $\varphi_e$ (that can be moved freely in some range and  probes the negative part of the potential) up to some radial  distance $\rho_\mm$.  
Beyond that core radius the field evolves smoothly towards the false vacuum $\varphi_+$. In the region outside the core the bounce equation (\ref{EoM4}) is satisfied while it is not inside the core and the correct value of $\rho_\mm$ decreases friction the right amount to satisfy $\varphi(\infty)=\varphi_\pp$. In spite of not being a proper solution of the bounce equation, the pseudo-bounce inherits some good properties of the proper bounces (although they do not extremize the Euclidean action). In particular, the slice of the pseudo-bounce 
at zero Euclidean time is a 3d bubble configuration with zero energy, and therefore the false vacuum can tunnel to it. This good property comes from the fact that the pseudo-bounce would exactly coincide with the proper bounce if the potential had a minimum at $\varphi_e$.

The picture for the decay of false vacua without bounce is that of a nearly flat direction (or valley) of the action functional in field configuration space, parametrized naturally by the value of $\varphi_e$. Among all configurations with fixed $\varphi_e$, pseudo-bounces have the lowest value of the tunneling action. 
Therefore, the field configurations along the bottom of that valley form a family of pseudo-bounce configurations. In a very concrete sense, pseudo-bounces are a particular instance of constrained instantons \cite{FY, Affleck}, which are an alternative route of attack for cases in which there is no bounce. Now, instead of an integral constraint on some field operator, pseudo-bounces
simply constrain the value of the field configuration at its core. This can also be achieved by an integral constraint with an operator of sufficiently high order that creates a minimum of $V(\varphi)$
at the $\varphi_e$ one wishes (see \cite{PseudoB} for details). 

The pseudo-bounces have different radii $R_e\equiv R(\varphi_e)$ and the rate of decay to a given pseudo-bounce would be
\be
\frac{\Gamma(\varphi_e)}{V}\simeq R_e^{-4} e^{-S(\varphi_e)}\ .
\ee
The total rate comes from summing over all pseudo-bounce decay channels
\be
\frac{\Gamma}{V}\simeq \int \frac{dR_e}{R_e}R_e^{-4} e^{-S(\varphi_e)}\ ,
\label{totalrate}
\ee
where the limits of integration are model dependent and eventually should regulate any possible divergence.

Even when there is a bounce, one can still consider what is the impact of decays to non-bounce field configurations on the total decay rate. When the bounce lies in a nearly flat direction in configuration space along which the action is not much bigger 
than the bounce one, then pseudo-bounces are again relevant, and it is important to integrate over them as in (\ref{totalrate}) to get a reliable estimate of the total rate. In some models it might happen that the integral (\ref{totalrate}) diverges (e.g. if decay configurations of arbitrarily small radius have a roughly constant tunneling action) signaling a catastrophically fast decay rate.

The problems of no bounce potentials and divergent decay rates
have been recently considered in a series of papers by Mukhanov and collaborators\cite{Mukhanov1,Mukhanov2,Mukhanov3,Mukhanov4}. In order to address and solve these problems they introduce a new type of
Euclidean configurations, dubbed new instantons. The main purpose of this paper is to confront their proposal with the existing solution based on pseudo-bounces.

The key idea of \cite{Mukhanov1,Mukhanov2,Mukhanov3,Mukhanov4} is to impose
 ultraviolet and infrared cutoffs on bounce field configurations. The argument used is that a classical field configuration, e.g. a bounce, is reliable only if the field value (and its derivatives) is larger than
 the expected quantum fluctuations. Decomposing the field configuration in momentum modes $\varphi_q$, the analysis of \cite{Mukhanov1,Mukhanov2,Mukhanov3,Mukhanov4} estimates the size of quantum fluctuations as $|\delta\varphi_q|\sim \sigma/\rho$ and $|\delta \dot\varphi_q|\sim \sigma/\rho^2$
with $\sigma$ an ${\cal O}(1)$ positive constant, which is the usual estimate when a (massless) quantum field is probed on a distance scale of size $\rho$. 

In practice, the value of $\dot\varphi(\rho)$ for a given bounce profile is compared to $\sigma/\rho^2$ and the profile is deemed to be classical (and trustable) for $\dot\varphi(\rho)>\sigma/\rho^2$
and quantum dominated (and not trustable) in the opposite case. Clearly, the stated  classicality condition is violated for $\rho<\rho_{uv}$ defined by
\be
\dot\varphi(\rho_{uv})=\frac{\sigma}{\rho_{uv}^2}\ ,
\label{rhouv}
\ee
(as the estimated size of quantum fluctuations diverges at $\rho\to 0$) as well as for $\rho>\rho_{ir}$ defined by
\be
\dot\varphi(\rho_{ir})=\frac{\sigma}{\rho_{ir}^2}\ ,
\label{rhoir}
\ee
(as the field derivative must go to zero faster than $1/\rho^2$ for $\rho\to\infty$ to have a finite action). 

Taking these cutoffs into account, Coleman's boundary condition
$\dot\varphi(0)=0$ (that fixes the bounce) is abandoned, and this allows for a whole family of field configurations: the new instantons. In practice, below $\rho_{uv}$ the new instanton profile is taken constant $\varphi(\rho)=\varphi_{uv}\equiv\varphi(\rho_{uv})$.
Doing the same for the infrared cutoff would give infinite Euclidean
action, so one should consider either $\varphi(\rho)=\varphi_\pp$
or the regular bounce profile for $\rho>\rho_{ir}$.

There are clear similarities between new instantons and pseudo-bounces. To confront and compare them, we reanalyze the potentials used in \cite{Mukhanov1,Mukhanov2} obtaining also the pseudo-bounce profiles and their Euclidean actions. In section~\ref{sec:linear}, we look at the linear unbounded potential of \cite{Mukhanov1}, case in which a potentially divergent decay rate is obtained as a result of small instanton contributions. Some of the particular lessons from this case regarding the comparison between actions and energies of new instantons and pseudo-bounces are of more general validity,  and this is proven in section~\ref{sec:General}.
The potential without bounce studied in \cite{Mukhanov2} is next examined
in section~\ref{sec:quartic} with the same comparative purpose. 
The final section~\ref{sec:discuss} contains a more in-depth
discussion about the physical implications of relying on pseudo-bounces or new instantons as well as on the physical basis on which both proposals stand. We advocate the use of pseudo-bounces and point out a number of shortcomings of the new instanton approach.

\section{Linear Unbounded Potential\label{sec:linear}}

In \cite{Mukhanov1}, the following potential is considered
(we follow the same notation to ease the comparison of results) 
\be
V(\varphi) = 
\begin{cases}
\lambda_\mm \varphi_0^3\varphi+ \displaystyle{\frac14}
\lambda_\pp \varphi_0^4\ , &  \mathrm{ for }\quad  \varphi<0\ ,
\\
\displaystyle{\frac14}\lambda_\pp (\varphi-\varphi^{}_0)^4\ , & \mathrm{ for } \quad \varphi>0\ .
 \end{cases}
 \label{Vlinear}
\ee
This potential has a false vacuum at $\varphi_0$ that can decay
towards the unbounded region for $\varphi<0$.
Following \cite{Mukhanov1}, $\lambda_\pp\ll 1$ is assumed 
so that loop corrections $\sim\lambda_\pp^2/(16\pi^2)$ can be neglected. On the other hand $\lambda_\mm$ can be taken to be sizable, leading to a very steep fall of the potential at $\varphi<0$.
  
In this potential the decay is governed by  a Coleman bounce with the profile
\be
\varphi(\rho)=\begin{cases}
\displaystyle{\frac18}\lambda_\mm\varphi_0^3(\rho^2-\rho_{0,C}^2)\ ,
 & \mathrm{for}\quad \rho<\rho_{0,C}\ ,\\
\varphi^{}_0\displaystyle{\frac{\rho^2-\rho_{0,C}^2}{\rho^2-\rho_{0,C}^2/(1+\delta_C)}}\ , & \mathrm{for}\quad \rho>\rho_{0,C}\ ,
\end{cases}
\label{CBL}
\ee
with
\be
\rho^2_{0,C}\equiv\frac{8}{\lambda_\mm \varphi_0^2(1-\beta)}\ ,\quad
\delta_C\equiv \frac{\lambda_\mm}{\lambda_\pp}\ ,
\label{paramsC}
\ee
where $\beta\equiv \lambda_\pp/(\lambda_\pp+\lambda_\mm)$. Here, $\rho_{0,C}$, the radius at which the field crosses zero, can be taken to be  the radius of the bounce.
The field value at the center of the bounce is
\be
\varphi(0)=
\varphi_{0,C}\equiv -\frac{\lambda_\pp+\lambda_\mm}{\lambda_\mm}\varphi_0\ .
\label{phiC}
\ee 
The tunneling action from this bounce reads
\be
S_C=\frac{8\pi^2}{3\lambda_\mm(1-\beta)^3}(2-\beta)(2-2\beta+\beta^2)\ .
\label{SC}
\ee

The analysis in \cite{Mukhanov1} takes issue with this result for $\lambda_\mm\gg 1$ (that gives $\beta\ll 1$) as, in that limit,
the (squared) radius of the bounce and the tunneling action can be made very small (both go as $1/\lambda_\mm$). It was argued in \cite{Mukhanov1}  that this would imply that the 
rate for decay is very fast, $\Gamma/V\sim \lambda_\mm^4 \exp[-32\pi^2/(3\lambda_\mm)]$, irrespective of the fact that the barrier height $V(0)=\lambda_\pp\varphi_0^4/4$ can be made very large (for large $\varphi_0$). 
However, the height of the potential barrier is not the key quantity here. To produce vacuum decay the field must pass configurations with $E>0$ tunneling through an energy barrier. A simple WKB analysis (see appendix for details) shows that, for the path in configuration space that minimizes the action, that energy barrier has height $\sim\varphi_0/\sqrt{\lambda_\mm}$ and width $\sim 1/(\varphi_0\lambda_\mm)$, leading to an action in exact agreement with (\ref{SC}) and simply explaining the parametric dependence $S_C\sim {\cal O}(100)/\lambda_\mm$.

\begin{figure}[t!]
\begin{center}
\includegraphics[width=0.45\textwidth]{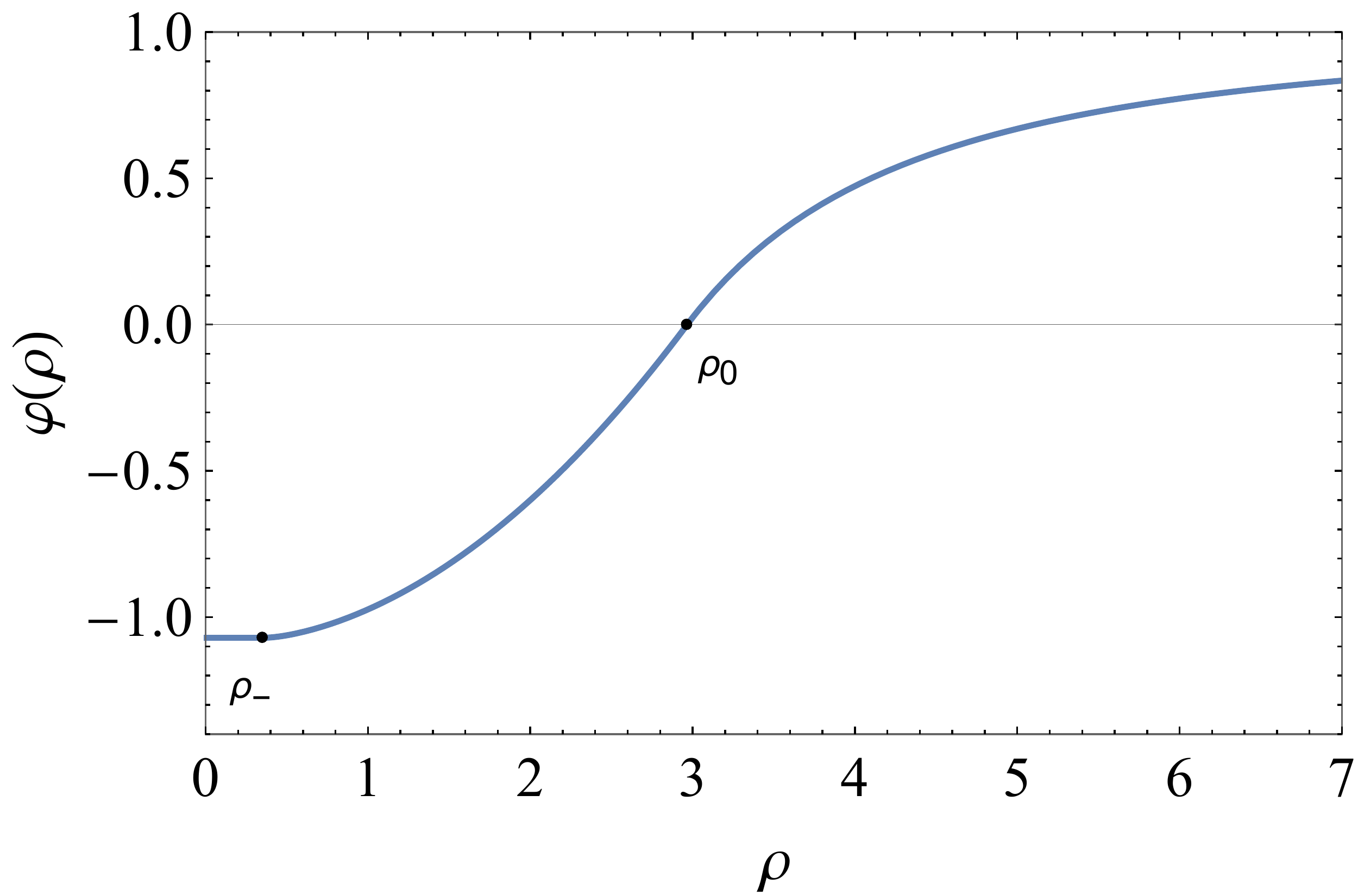}
\hspace{0.2cm}
\includegraphics[width=0.45\textwidth]{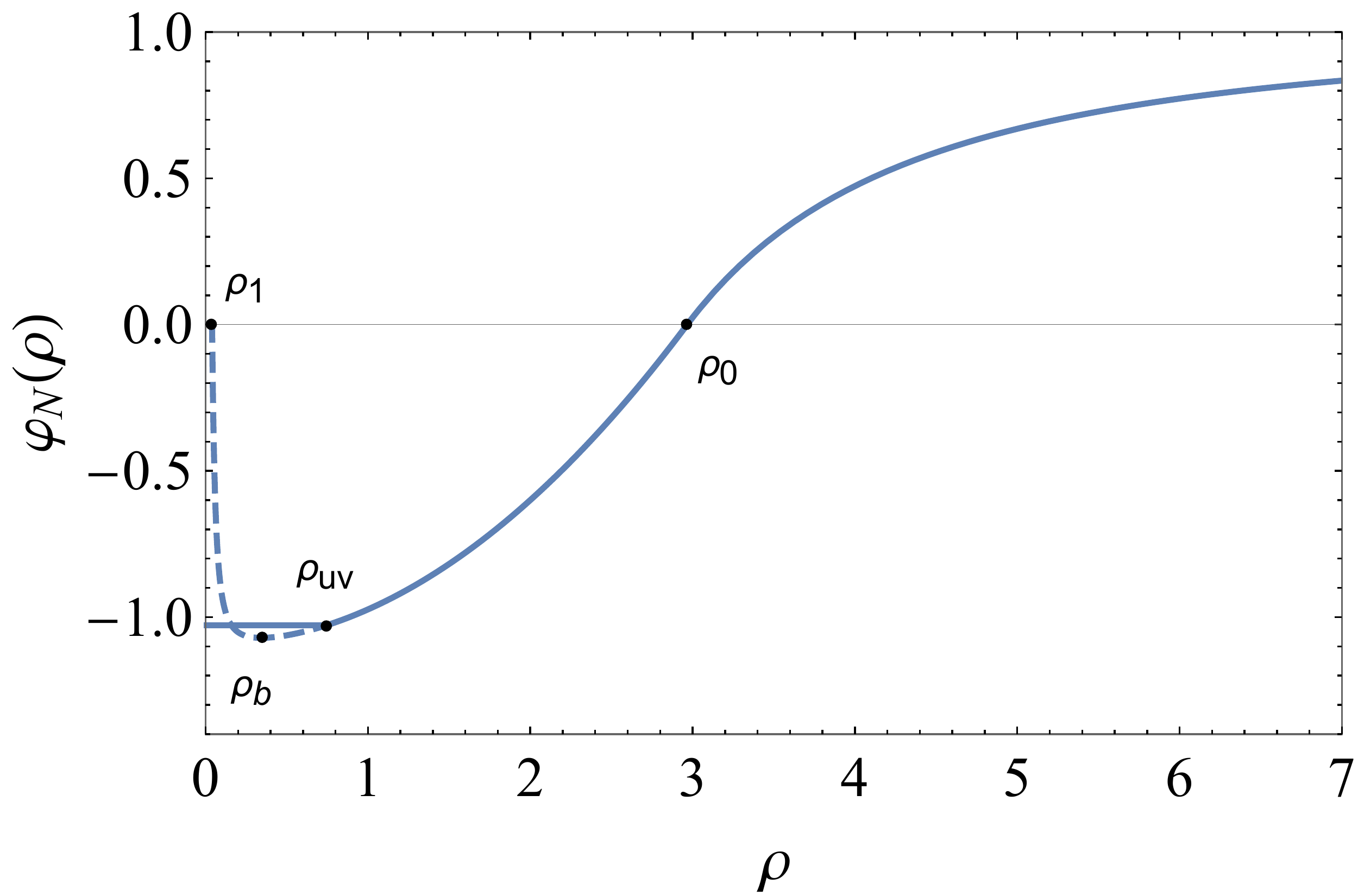}
\end{center}
\caption{ Field profiles for pseudo-bounce (left) and new instanton (right) for the unbounded linear potential of (\ref{Vlinear}) with $\lambda_\mm=1$, $\lambda_\pp=0.1$, $\varphi_0=1$ and $\rho_0=2.97$ (leading to $\rho_1=0.04$,  $\rho_\mm=\rho_b=0.34$, $\rho_{uv}=0.75$).
\label{fig:Profiles}
}
\end{figure}

In any case, the previous issue was identified by \cite{Mukhanov1} as an ultraviolet problem that needed to be fixed and motivated the introduction of new instantons, in which the presence of an ultraviolet cutoff $\rho_{uv}$ on the bounce profile could fix this problem. Before describing how this works out, let us first consider pseudo-bounces. Even though there is a bounce for this potential, the false vacuum can still decay to other field configurations like pseudo-bounces. If their action is comparable to the bounce one, one needs to sum over them to evaluate the decay rate accurately \cite{PseudoB}.
The pseudo-bounce profile for this potential, according to the prescription detailed in \cite{PseudoB}, is
\be
\varphi(\rho)=\begin{cases}
\varphi_{0\mm}^{}\ , & \mathrm{for}\quad \rho<\rho_\mm\ ,
\\
\displaystyle{\frac18}\lambda_\mm\varphi_0^3(\rho^2-\rho_0^2)\left(1-\displaystyle{\frac{\rho_\mm^4}{\rho^2\rho_0^2}}\right)\ ,
 & \mathrm{for}\quad \rho_\mm<\rho<\rho_0\ ,\\
\varphi^{}_0\displaystyle{\frac{\rho^2-\rho_0^2}{\rho^2-\rho_0^2/(1+\delta)}}\ , & \mathrm{for}\quad \rho>\rho_0\ ,
\end{cases}
\ee
where
\bea
\delta &=& \frac{4}{\lambda_\pp\varphi_0^2\rho_0^2}\left(1+\sqrt{1+\lambda_\pp\varphi_0^2\rho_0^2/2}\right)\ ,\\
\rho_\mm^4 &=&\rho_0^4-\left(1+\frac{1}{\delta}\right)\frac{8\rho_0^2}{\lambda_\mm\varphi_0^2}\ ,\\
\varphi_{0\mm}&=&-\frac{\lambda_\mm\varphi_0^3}{8\rho_0^2}\left(\rho_\mm^2-\rho_0^2\right)^2\ , 
\eea
so that $\varphi(\rho)$ and its derivative are continuous at $\rho_\mm$ and $\rho_0$. The pseudo-bounce is constant for $\rho<\rho_\mm$ and is a solution of the bounce equation (\ref{EoM4}) in the range $\rho>\rho_\mm$.
This gives a one-parameter family of pseudo-bounces that can be parametrized either by $\rho_0$, $\rho_\mm$ or $\varphi_{0\mm}$\footnote{We only consider pseudo-bounces with $|\varphi_{0\mm}|<|\varphi_{0,C}|$. For $|\varphi_{0\mm}|>|\varphi_{0,C}|$, the field configuration that minimizes the tunneling action is singular, identical to the Coleman bounce except for a spike at $\rho=0$, with the field jumping to $\varphi_{0\mm}$ there and action equal to the Coleman one. We discard such configurations.} and includes the as a limit case the proper bounce, that corresponds to $\rho_\mm=0$.
The field profile of one pseudo-bounce is shown in the left plot of figure~\ref{fig:Profiles}.

Concerning new instantons, their key feature is the presence of a ``quantum core''
below $\rho_{uv}$, defined in (\ref{rhouv}), which in this model gives the equation
\be
\frac14\lambda_\mm\varphi_0^3\left(\rho_{uv}^4-\rho_b^4\right)=\sigma \rho_{uv}\ ,
\label{rhouvl}
\ee
with $\rho_b\equiv\rho_\mm$.
Nevertheless, the field profile of new instantons, $\varphi_N(\rho)$, is in fact rather similar to the pseudo-bounce profile just discussed.
For $\rho>\rho_{uv}$ both profiles are exactly the same  (as both are solutions of the Euler-Lagrange equation for the bounce in that region). They only depart below $\rho_{uv}$, where the new instanton is assumed to be dominated by quantum fluctuations. In practice, for calculating the action or the energy,  the field is assumed to be $\varphi_{uv}=\varphi_N(\rho_{uv})$ in that core region $\rho<\rho_{uv}$. In contrast, the pseudo-bounce reaches down to $\rho_\mm=\rho_b$ (where $\dot\varphi=0$) and then stays flat with $\varphi=\varphi_{0\mm}=\varphi(\rho_\mm)$. See right plot of Fig.~\ref{fig:Profiles} for an example of new instanton profile to be compared with the pseudo-bounce profile on the left, for the same value of $\rho_0$, where the field crosses zero. The dashed line extending below $\rho_{uv}$ shows the singular solution to the bounce equation, with the locations of $\rho_b$ (where $\dot\varphi=0$)
and $\rho_1$ (where $\varphi=0$) indicated by black dots. 

Once the profiles are known, the Euclidean action for decay via these configurations can be calculated analytically. For pseudo-bounces one gets
\bea
S&=&\frac{\pi^2}{192\lambda_\pp}\left[-256+6K^2(2\delta_C-1)-K^3\delta_C^2
+2(64-8K+3K^2\delta_C)\sqrt{4+K}\right.\nonumber\\
&&\left. +\sqrt{\delta_C}\left(K^2\delta_C-8K-4K\sqrt{4+K}\right)^{3/2}\right]\ ,
\eea
where $K\equiv 2\lambda_\pp\varphi_0^2\rho_0^2$ and $\delta_C$ is defined in (\ref{paramsC}).
The action $S_N$ for the new instantons is calculated in \cite{Mukhanov1}. 

\begin{figure}[t!]
\begin{center}
\includegraphics[width=0.5\textwidth]{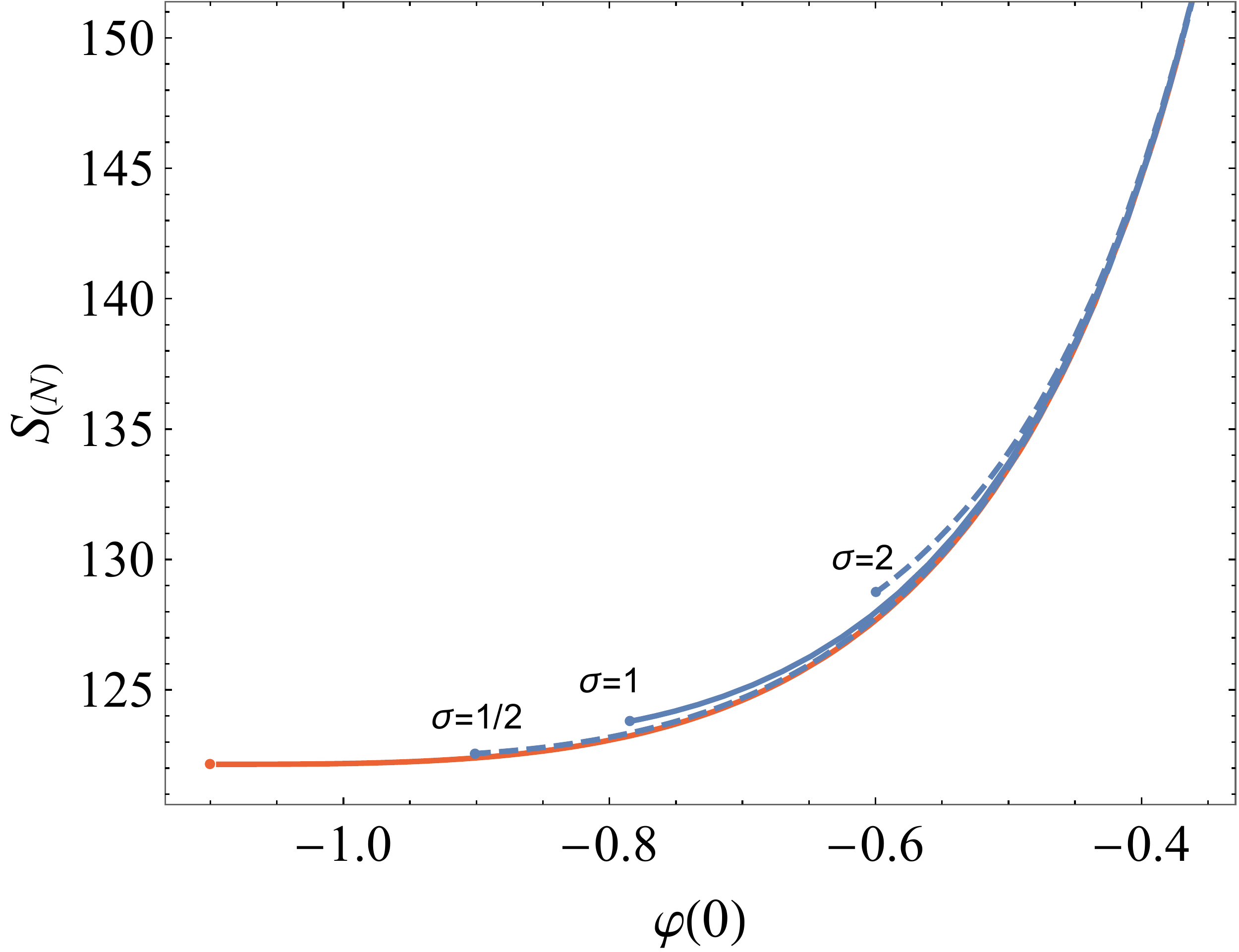}
\end{center}
\caption{Decay action  via pseudo-bounces (red solid line) or new instantons (blue lines, for three values of $\sigma$ as indicated) as a function of the core value of the field configuration mediating decay, for the potential (\ref{Vlinear}) with $\lambda_\mm=1$, $\lambda_\pp=0.1$, $\varphi_0=1$.
\label{fig:actions}
}
\end{figure}

Figure~\ref{fig:actions} compares the action for the pseudo-bounces
(red line) with those for the new instantons (for $\sigma=1$, solid line and $\sigma=1/2,2$, dashed lines), as a function of $\varphi(0)$, the value of $\varphi$  at its center (this is $\varphi_{0\mm}$ for pseudo-bounces and $\varphi_{uv}$ for new instantons). The action for pseudo-bounces is minimized when $\varphi(0)$ is equal to the value for the Coleman bounce, at which point the pseudo-bounce tends to the Coleman bounce. We also see that the action of  pseudo-bounces is smaller than that of new instantons. This is expected, as the pseudo-bounce minimizes the Euclidean action for fixed $\varphi(0)$, see \cite{PseudoB}. The difference between the actions is not large and is within the uncertainty one could associate with quantum fluctuations over the semiclassical result, but grows with increasing $\sigma$. For small $\sigma$, the new instanton profile tends to be degenerate with the pseudo-bounce (as $\rho_{uv}\to\rho_b$ for $\sigma\to 0$) and the difference between the corresponding actions can be obtained analytically in that limit as
\be
S_N-S=\frac14\pi^2\sigma^2 +{\cal O}(\sigma^3)\ .
\ee
This simple result is in fact general, as proven in the next section.

\begin{figure}[t!]
\begin{center}
\includegraphics[width=0.45\textwidth]{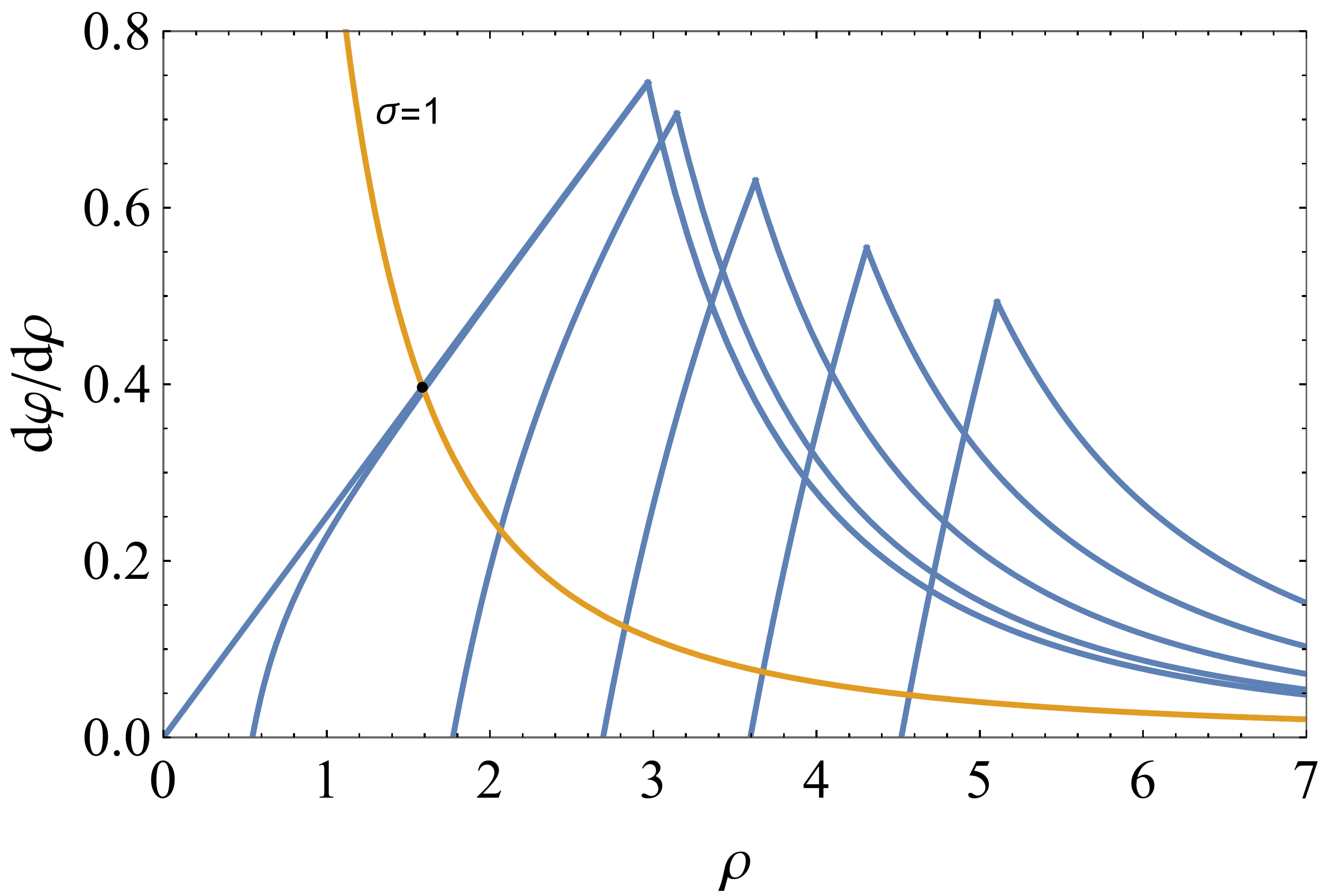}
\hspace{0.2cm}
\includegraphics[width=0.475\textwidth]{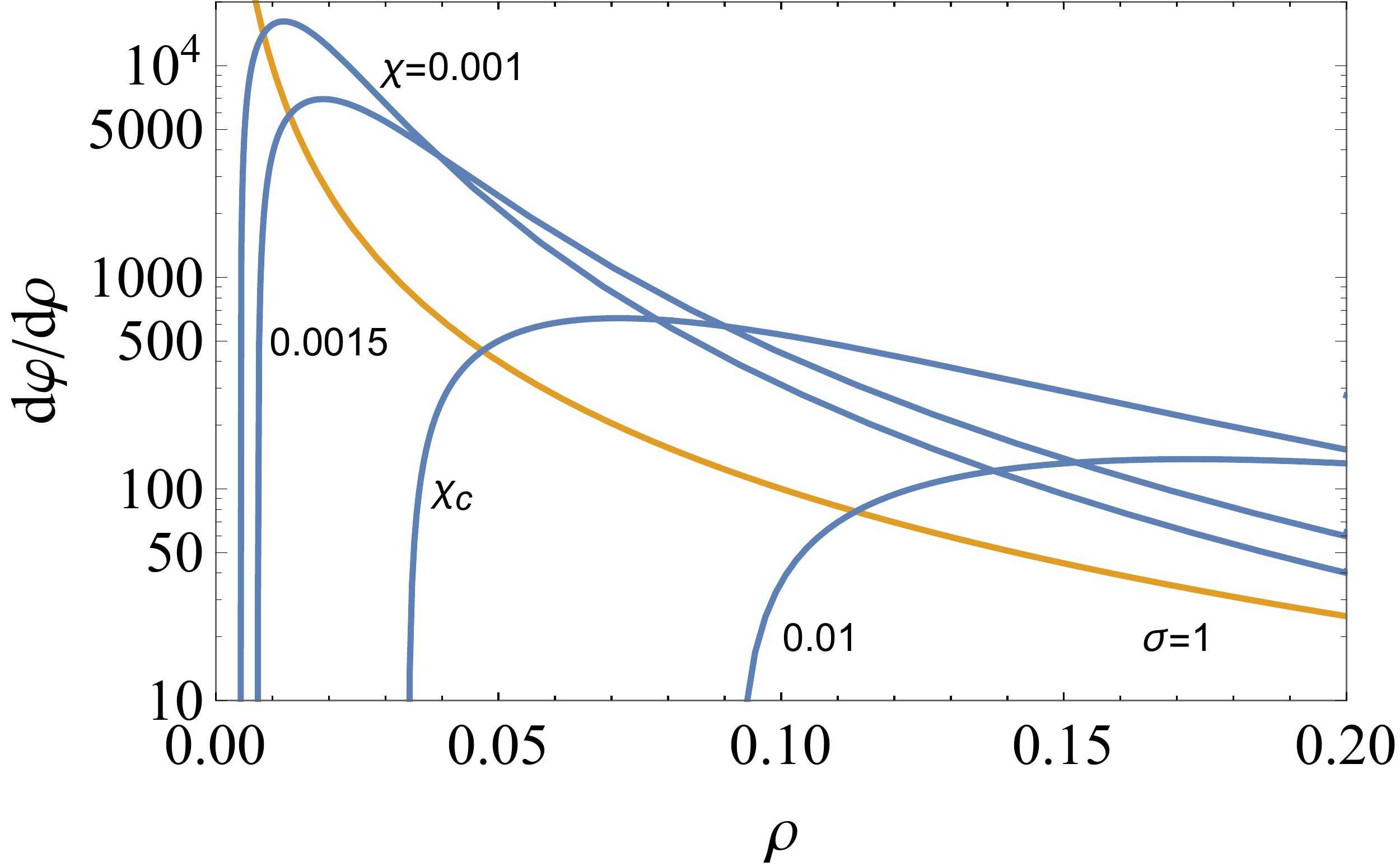}
\end{center}
\caption{Slope $d\varphi/d\rho$ for pseudo-bounces/new instantons. Left plot: for the linear potential (\ref{Vlinear}) and several values of $\rho_\mm$, with $\lambda_\mm=1$, $\lambda_\pp=0.1$, $\varphi_0=1$. Right plot: same for the quartic potential (\ref{Vquartic}) and several values of $\chi$, with 
$\lambda_\mm=\lambda_\pp=0.1$, $\varphi_0=1$.  The ``quantum slope'' $\sigma/\rho^2$ is also plotted, for $\sigma=1$.
\label{fig:CoreLimit}
}
\end{figure}

The difference between both actions is larger if we compare  the range of $\varphi(0)$ that is allowed. For the new instantons there is a lower limit on $\varphi(0)$ that is reached when $\rho_b\to 0$, in which case $\rho_{uv}$ takes its minimal value 
\be
\mathrm{Min}[\rho_{uv}]=\frac{1}{\varphi_0}
\left(\frac{4\sigma}{\lambda_\mm}\right)^{1/3}\ ,
\ee
as follows from (\ref{rhouvl}). This leads to a minimal value
for $\varphi(0)$, $ \mathrm{Min}[\varphi(0)]=\varphi(\mathrm{Min}[\rho_{uv}])$. Using $\rho_0^2\varphi_0^2=8(\lambda_\pp+\lambda_\mm)/\lambda_\mm^2$, we find
\be
\mathrm{Min}[\varphi(0)]= \varphi_{0,C}+\frac14\varphi_0\left(2\lambda_\mm\sigma^2\right)^{1/3}\ ,
\ee
where $\varphi_{0,C}$  is $\varphi(0)$ for the Coleman bounce, see (\ref{phiC}). This behaviour is illustrated by the left plot of figure~\ref{fig:CoreLimit} which shows $\dot\varphi$ for pseudo-bounces (blue lines) for several values of $\rho_\mm$ (given by the intersection of the lines with the $\rho$-axis) and the curve $\sigma/\rho^2$ for $\sigma=1$ (labeled $\sigma=1$), which intersects the blue lines at $\rho_{uv}$. For new instantons, $\dot\varphi$ drops to zero below $\rho_{uv}$, while it is the same as for pseudo-bounces above it. The point for $\mathrm{Min}[\rho_{uv}]$ is marked by a black dot.

We see that the difference between the minimum value of $\varphi(0)$ 
 for the new instanton and the Coleman value is not necessarily small, as is also illustrated in figure~\ref{fig:actions} where the lowest point of each curve corresponds to $\mathrm{Min}[\varphi(0)]$. This difference can be relevant as it implies a sizable difference in the minimum allowed value of the tunneling action, leading to order-of-magnitude differences in the decay rate.

As explained in \cite{PseudoB}, the energy of the critical bubble associated to the pseudo-bounce (the 3d slice at zero Euclidean time, $\tau=0$, of the 4D Euclidean pseudo-bounce) is exactly zero. Notice that the same energy for the new instanton does not vanish, but is $2\pi\sigma^2/(3\rho_{uv})$ \cite{Mukhanov1}, see next section.

\section{Some General Results\label{sec:General}}

The key difference between pseudo-bounce and new instanton profiles is due to the difference between $\rho_{uv}$ and $\rho_\mm=\rho_b$,  which is ${\cal O}(\sigma)$. By expanding $\dot\varphi_{uv}\equiv \dot\varphi(\rho_{uv})$ around $\rho_b$, one gets
\be
\dot\varphi_{uv} \equiv \frac{\sigma}{\rho_{uv}^2}= \dot\varphi_b +\ddot\varphi_b\delta\rho + {\cal O}(\delta\rho^2)\ ,
\label{deltarho0}
\ee
where $\delta\rho\equiv\rho_{uv}-\rho_b$. By definition, $\dot\varphi_b\equiv\dot\varphi(\rho_b)=0$, and therefore the bounce equation (\ref{EoM4}) gives $\ddot\varphi_b=V'_b$. Using these results in (\ref{deltarho0}) we get
\be
\delta\rho=\frac{\sigma}{V_b'\rho_b^2}+{\cal O}(\sigma^2)\ .
\label{deltarho}
\ee
The difference between the actions for new instanton, $S_N$, and pseudo-bounce, $S$, is 
\be
\Delta S = S_N - S =-2\pi^2\int_{\rho_{b}}^{\rho_{uv}}d\rho\rho^3\left[\frac12 \dot\varphi^2+V(\varphi)\right]+\frac{\pi^2}{2}\left(\rho_{uv}^4V_{uv}-\rho_b^4V_b\right)\ .
\ee
Expanding this difference up to second order in $\delta\rho$ gives
\be
\Delta S = \frac{\pi^2}{4}\rho_b^4V_b'\ddot\varphi_b \delta\rho^2+{\cal O}(\delta\rho^3)\ ,
\ee
and using (\ref{deltarho}) 
\be
\Delta S =\frac14 \pi^2\sigma^2 + {\cal O}(\sigma^3)\ .
\ee
So that this is indeed a general, potential-independent result.

A similar comparison can be made for the total energy of the 
3d tunneling bubble configuration (the $\tau=0$ slice of the 4d Euclidean configuration). As proven in \cite{PseudoB}, pseudo-bounces automatically lead to $E=0$ tunneling bubbles, while for 
new instantons we get
\be
\Delta E\equiv E_N-E=-4\pi\int_{\rho_{b}}^{\rho_{uv}}d\rho\rho^2\left[\frac12 \dot\varphi^2+V(\varphi)\right]+\frac{4\pi}{3}\left(\rho_{uv}^3V_{uv}-\rho_b^3V_b\right)\ ,
\ee
and using the results above for $\ddot\varphi_b$ and $\delta\rho$
\be
\Delta E = E_N = \frac{2\pi\sigma^2}{3\rho_b}+{\cal O}(\sigma^3)\ ,
\ee
as found in \cite{Mukhanov1}. 
  
\section{Quartic Unbounded Potential\label{sec:quartic}}

In this section we examine one potential without bounce (or rather with bounce pushed away to infinity), of the type described in section~\ref{sec:intro}, that was studied in \cite{Mukhanov2}. 
The potential is:
\begin{equation}
    V(\varphi)=
    \begin{cases}
      -\displaystyle{\frac{\lambda_-}{4}}(\varphi^4-\beta^3\varphi_0^4)\ , & \text{for}\ \varphi< \beta\varphi_0 \\
     \displaystyle{ \frac{\lambda_+}{4}}(\varphi-\varphi_0)^4\ , & \text{for}\ \varphi> \beta\varphi_0
    \end{cases}
    \label{Vquartic}
\end{equation}
where $\lambda_\pm>0$, 
$\beta=\lambda_+^{1/3}/(\lambda_+^{1/3}+\lambda_-^{1/3})$ and we take $\varphi_0>0$.

This unbounded-from-below potential features an unstable false vacuum at $\varphi_0$ but there is no Coleman bounce. This should be clear from our discussion in section~\ref{sec:intro} and can be checked explicitly by using the general formula (\ref{BCond}), which leads to the condition that the bounce should satisfy
\be
\frac14\lambda_\mm\beta^3\rho_m^4+\lambda_\pp\int_{\rho_m}^\infty (1-\varphi/\varphi_0)^3\rho^3d\rho =0\ ,
\label{nob}
\ee
where $\rho_m$ is defined by $\varphi(\rho_m)=\beta\varphi_0$.
As $\beta\varphi_0 <\varphi<\varphi_0$ in the range $(\rho_m,\infty)$, the left hand side of (\ref{nob}) is positive definite and no bounce can exist.

This is one more example of tunneling without bounce studied in \cite{PseudoB} and leading to the introduction of pseudo-bounces.
The pseudo-bounce profile in this particular case follows the general structure described in \cite{PseudoB} and section~\ref{sec:intro}: it has a core with constant field value below some $\rho_\mm$ and, above that value, the profile is a solution of the Euclidean equation of motion that tends to $\varphi_0$ at $\rho\to\infty$. Using notation from \cite{Mukhanov2} to ease the comparison with new instantons, the pseudo-bounce profile has the form
\be
    \varphi(\rho)=
    \begin{cases}
    \varphi_{0_-}\ , & \text{for}\ \rho<\rho_- \\
      \frac{\sqrt{2}k}{\rho\sqrt{\lambda_-(2k^2-1)}}\text{cn}\left(\frac{\ln(\rho/\alpha)}{\sqrt{2k^2-1}},k\right)\ , & \text{for}\ \rho_-<\rho< \rho_m \\
     \varphi_0\frac{\rho^2-\rho_0^2}{\rho^2-\rho_0^2/(1+\delta)}\ , & \text{for}\ \rho> \rho_m
    \end{cases}
\ee
where $\rho_0$ and $\alpha$ are integration constants, $cn(u,k)$ is the Jacobi elliptic function\footnote{For $u=\int_0^\phi d\theta/\sqrt{1-k^2 \sin^2\theta}$, $cn(u,k)=\cos\phi$, $sn(u,k)=\sin\phi$ and $dn(u,k)=\sqrt{1-k^2\sin^2\phi}$.}, $\varphi_m\equiv\varphi(\rho_m)=\beta\varphi_0$ and $\varphi_{0_-}\equiv \varphi(\rho_-)$. Instead of $\rho_0$, it is convenient to introduce the variable $\chi$ by \cite{Mukhanov2}
\be
\rho_0^2\varphi_0^2 = \frac{E_c}{4(1-\beta)}\chi(1-\beta+\chi)\ ,
\ee 
where 
\be
E_c\equiv\frac{32}{\lambda_\pp(1-\beta)}\ .
\ee
We then have
\be
\delta =\frac{1-\beta}{\chi}\ ,\quad\quad \rho_m^2\varphi_0^2=\frac{E_c}{4(1-\beta)}\chi(1+\chi)\ ,
\ee
and
\be
k =\frac{1}{\sqrt{2}}\sqrt{1+\frac{1}{\sqrt{1+2\lambda_-E_-}}}\ , 
\ee
with  
\be
E_\mm =E_c\chi(1+\chi)^3\ .
\ee
Note that $1/\sqrt{2}<k< 1$.
At $\rho_\mm$ one has $\dot\varphi=0$ so that $\rho_\mm$ is equal to the $\rho_b$ of \cite{Mukhanov2}. The core field value $\varphi_{0\mm}$ is free and can be used to parametrize the family of pseudo-bounces. Alternatively one can use the constant $\rho_0$
or the variable $\chi$ introduced above.

The constants $\alpha$ and $\rho_-$ are determined by imposing the continuity of $\varphi(\rho)$ at the matching radii $\rho_m$ and $\rho_\mm$:
\be
\text{cn}\left(\frac{\ln(\rho_m/\alpha)}{\sqrt{2k^2-1}},k\right)=\sqrt{\lambda_-\beta^2\rho_m^2\varphi_0^2\left(1-\frac{1}{2k^2}\right)}\equiv \epsilon_m>0\ ,\ \ \  \text{sn}\left(\frac{\ln(\rho_m/\alpha)}{\sqrt{2k^2-1}},k\right)<0 \ .
\ee

\be
\text{cn}\left(\frac{\ln(\rho_-/\alpha)}{\sqrt{2k^2-1}},k\right)=-\left(\frac{1}{k^2}-1\right)^{1/4}\equiv -\epsilon_b<0\ ,\ \ \  \text{sn}\left(\frac{\ln(\rho_-/\alpha)}{\sqrt{2k^2-1}},k\right)>0
\ ,
\ee
with the sign of $sn(u,k)$ chosen to get $\dot\varphi(\rho_m)>0$
and $\dot\varphi(\rho_\mm)=0$.

As for the potential in the previous section, the field profile of the new instanton (described in detail in \cite{Mukhanov2}) coincides with the pseudo-bounce profile except for the fact that the constant-field core extends up to $\rho_{uv}>\rho_\mm=\rho_b$. We do not present a figure comparing the two profiles in this case as it would be quite similar to Fig.~\ref{fig:Profiles}.

 For the potential at hand, the condition (\ref{rhouv}) defining $\rho_{uv}$, takes the form \cite{Mukhanov2}
\be
\left(\frac{\epsilon_{uv}}{\epsilon_b}\right)^4-
\left(\frac{32\sigma^4}{\lambda_\mm E_\mm^3}\right)^{1/4}\frac{\epsilon_{uv}}{\epsilon_b}
-\left(1-\frac{\sigma^2}{E_\mm}\right)=0\ ,
\label{epsilonuv}
\ee
where $\epsilon_{uv}$ is related to $\rho_{uv}$ by
\be
\text{cn}\left(\frac{\ln(\rho_{uv}/\alpha)}{\sqrt{2k^2-1}},k\right)\equiv -\epsilon_{uv}<0\ ,\ \ \  \text{sn}\left(\frac{\ln(\rho_{uv}/\alpha)}{\sqrt{2k^2-1}},k\right)>0
\ ,
\ee
with the sign of $sn(u,k)$ fixed to get $\dot\varphi(\rho_{uv})>0$.
Solving (\ref{epsilonuv}), one gets
\be
\frac{\epsilon_{uv}}{\epsilon_b}=\frac{1}{3^{1/4}}\left[\sqrt{W_N}+\sqrt{\left(\frac{\gamma_N}{W_N}\right)^{1/2}-W_N}\right]\ ,
\ee    
with
\bea
W_N &\equiv &\frac12\left[\left(\gamma_N+\sqrt{z^3+\gamma^2_N}\right)^{1/3}-z\left(\gamma_N+\sqrt{z^3+\gamma^2_N}\right)^{-1/3}\right]\ ,\\
\gamma_N &\equiv &  
\sqrt{
\frac{27\sigma^4}{8\lambda_\mm E_\mm^3}
}\ ,\quad\quad
z\equiv  1-\frac{\sigma^2}{E_\mm}\ .
\eea   
Note that this solution differs from the one presented in \cite{Mukhanov2}.

Once the field profiles are known, it is a simple matter to obtain the corresponding tunneling actions.
The Euclidean action $S$ for the pseudo-bounces can be calculated analytically and reads
\bea
        S&= & \frac{\pi^2}{3}\left\{\frac{E_-}{4}\left[\frac{1}{1+\chi}+\frac{1-2\beta}{(1-\beta)(1+\chi)^2}+\frac{1-\beta}{(1+\chi)^3}\right]+\sqrt{\frac{2E_-}{\lambda_-}}+\frac{E_-(2-3k^2)}{k^2}\ln\left(\frac{\rho_m}{\rho_\mm}\right)\right. \nonumber\\
        &&+\left. \frac{4(1+2\lambda_-E_-)^{1/4}}{\lambda_-}\left[4\textbf{E}(k)-\textbf{E}(\arccos(\epsilon_m),k)-\textbf{E}(\arccos(-\epsilon_b),k)\right] \right\}\ .
\label{Sq}        
\eea
Here $\textbf{E}(k)$ and $\textbf{E}(\phi,k)$ are, respectively, the complete and incomplete elliptic integrals of the second kind:
\be
\textbf{E}(\phi,k)\equiv \int_0^\phi\sqrt{1-k^2\sin^2\theta}\, d\theta\ ,\quad
\textbf{E}(k)\equiv \textbf{E}(\pi/2,k)\ .
\ee

The Euclidean action for the new instantons, $S_N$, is slightly more complicated:
\bea
        S_N&=& \frac{\pi^2}{3}\left\{\frac{E_-}{4}\left[3+\frac{1}{1+\chi}+\frac{1-2\beta}{(1-\beta)(1+\chi)^2}+\frac{1-\beta}{(1+\chi)^3}\right]+\frac{E_-(2-3k^2)}{k^2}\ln\left(\frac{\rho_m}{\rho_\mm}\right)\right. \nonumber\\
       & &+ \frac{4(1+2\lambda_-E_-)^{1/4}}{\lambda_-}\left[4\textbf{E}(k)-\textbf{E}(\arccos(\epsilon_m),k)-\textbf{E}(\arccos(-\epsilon_{uv}),k)\right]  \nonumber\\
   && \left. +2\sigma\left(\frac{2E_-}{\lambda_-}\right)^{1/4}\frac{\epsilon_{uv}}{\epsilon_b}  +\sqrt{\frac{2E_-}{\lambda_-}}\left(\frac{\epsilon_{uv}}{\epsilon_b}\right)^2-\frac{3}{4}E_\mm
     \left(\frac{\epsilon_{uv}}{\epsilon_b}\right)^4   \right\}\ ,
\label{SNq}
\eea
with the last term missing in the result obtained in \cite{Mukhanov2}.

\begin{figure}[t!]
\begin{center}
\includegraphics[width=0.5\textwidth]{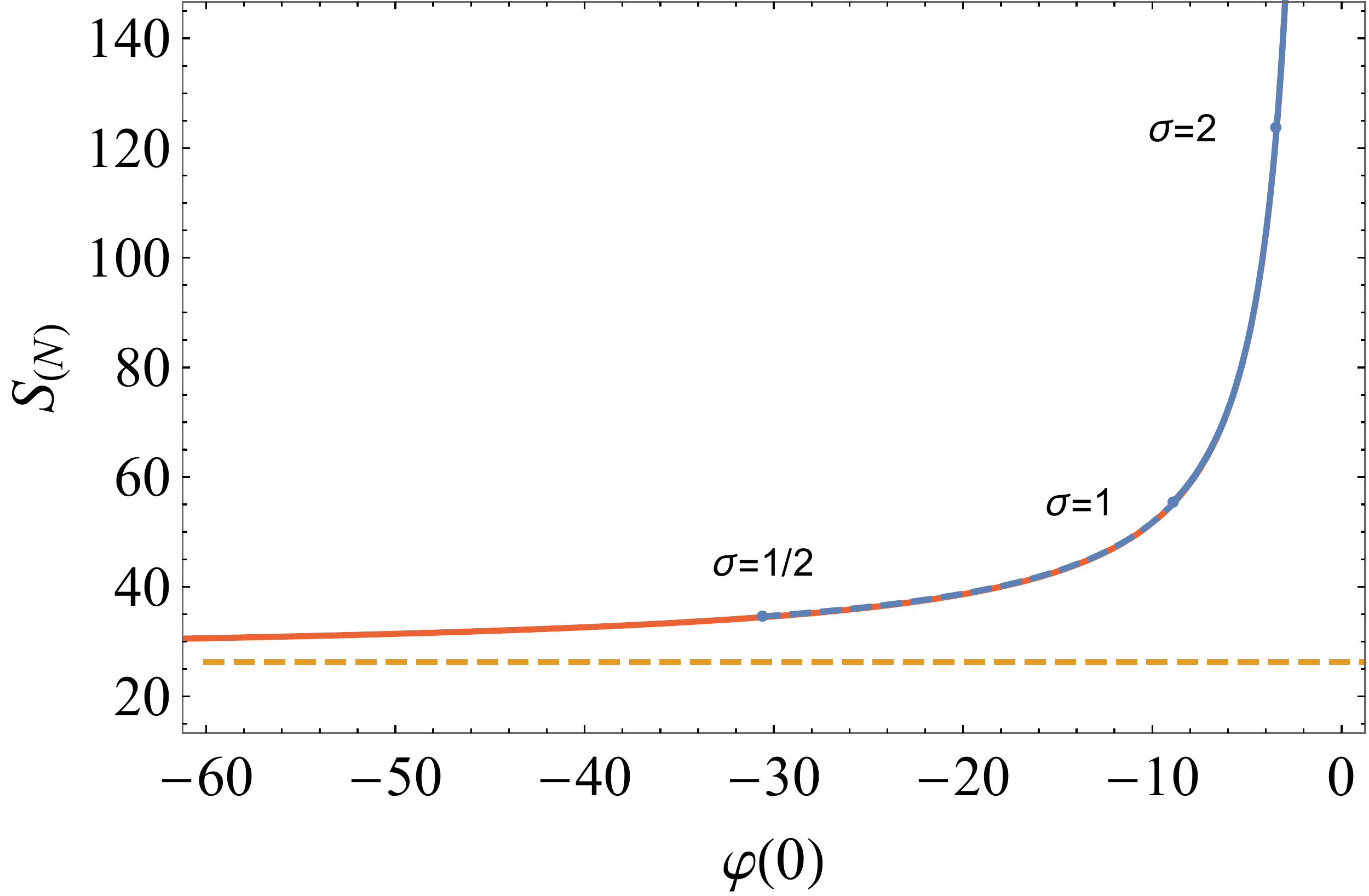}
\end{center}
\caption{Decay action  via pseudo-bounces (red solid line) or new instantons (blue lines, for three values of $\sigma$ as indicated) as a function of the core value of the field configuration mediating decay, for the potential (\ref{Vquartic}) with $\lambda_\mm=1$, $\lambda_\pp=0.1$ and $\varphi_0=1$. The dashed line is the asymptotic value (\ref{Sasympt}).
\label{fig:actionsquartic}
}
\end{figure}

In Fig.~\ref{fig:actionsquartic} we compare both actions, with $S$ in red and $S_N$ in blue for $\sigma=1$ (solid line) and $\sigma=1/2,2$ (dashed lines) for $\lambda_\mm=1$, $\lambda_\pp=0.1$ and $\varphi_0=1$. Asymptotically,
as $\varphi(0)\to-\infty$, the pseudo-bounce action (\ref{Sq}) tends to the standard value for a negative quartic potential
\be
S\to\frac{8\pi^2}{3\lambda_\mm}\ ,
\label{Sasympt}
\ee
indicated in Fig.~\ref{fig:actionsquartic} with a horizontal dashed line. This is expected: when $\varphi(0)$ is much larger than any other mass scale in the problem, the details of the potential barrier
are unimportant and the pseudo-bounce tends to the Fubini bounce for a $V=-\lambda_\mm\varphi^4/4$ potential, with action given by (\ref{Sasympt}).

As shown in the previous section on general grounds and illustrated by figure~\ref{fig:actionsquartic} , the new instanton action, $S_N$ is bigger than $S$ but close to it. The main difference regarding the final conclusion for the decay rate is that the new instantons
have a lower limit on $\varphi(0)$, indicated by the dots in figure~\ref{fig:actionsquartic} and, therefore, $S_N$ can be significantly bigger than the asymptotic value of $S$. In addition, due to this effect, the uncertainty in the parameter $\sigma$ results in a significant uncertainty in the new instanton action (and thus the rate). The lower bound on $\varphi(0)$ comes from the constraint $E_\mm > 3\sigma^2$
estimated in \cite{Mukhanov2}, that can be rewritten as a lower limit on $\chi$ as
\be
\chi>\chi_c \equiv \frac{3\sigma^2}{E_c}\ .
\ee
We find that this condition can be violated. The right plot of fig.~\ref{fig:CoreLimit} shows $\dot\varphi$ of new instantons/pseudo-bounces for several values of $\chi$ as well as the quantum limit $\sigma/\rho^2$ for $\sigma=1$. For the chosen parameters, it is shown that values $\chi<\chi_c$ are still possible, and in fact there seems to be no lower limit for $\chi$. For other parameter choices, imposing the quantum condition would result into a lower limit on $\varphi(0)$ but we do not pursue this matter further.

The final decay rate in this model results from integrating along the flat valley shown in fig.~\ref{fig:actionsquartic}, integral that might diverge, depending on the ultraviolet physics that eventually stabilizes the potential. New instantons would offer a solution to this divergence if there was a finite $\mathrm{Min}[\varphi(0)]$. As we have seen, in some cases no such limit exists, and even when it does, see next section, there is no true obstruction to reaching values below that limit.

\section{Discussion \label{sec:discuss}}

{\bf $\bma{ \S\, 1}$ Final states of the vacuum decay process.}
Consider a scalar potential that has a local minimum at some $\varphi_0$ and, in the absence of gravity, set $V(\varphi_0)=0$ without loss of generality. 
The decay of such false vacuum towards a deeper one is a tunneling process between two field configurations: the initial one is homogeneous, with $\varphi(\vec{x})=\varphi_0$, while the final one is some bubble-like field configuration $\varphi_b(\vec{x})$. The latter configuration must probe the negative part of the  potential such that, once produced, it will grow and convert all space to the deeper vacuum phase. It should also have zero energy (to conserve energy in the tunneling process) and this is achieved by a compensation between the negative energy in its bulk and the positive energy of the field gradient in its wall.

The space of possible $\varphi_b(\vec{x})$ configurations with $E=0$ is infinite dimensional\footnote{In the presence of strong gravitational effects, such $E=0$ configurations in AdS or Minkowski false vacua might not exist. When that happens, there is nowhere to tunnel to  and gravity stabilizes the false vacuum \cite{CdL}. (For a recent simple description, see \cite{Estab})} and in principle nothing  prevents the false vacuum from decaying into any one of these configurations.  A complete calculation of the decay rate should sum over all possible final states of the decay process.  When there exists a (4d Euclidean) Coleman bounce, its
$\tau=0$ slice gives a (spherically symmetric) bubble, $\varphi_B(r)$, singled out among all $E=0$ configurations by the fact that it has the minimum Euclidean action\footnote{The bounce is a saddle point of the Euclidean action but is a minimum when only $E=0$ configurations are considered \cite{CN}. (See also \cite{Rubakov} for an excellent introduction to these topics.)} and is therefore the most likely outcome of the decay process and the one that dominates in the sum over final states.

Pseudo-bounces have $E=0$ and are therefore possible final states of false vacuum decay.  New instantons in principle do not have $E=0$, as shown in section~\ref{sec:General}, although they fail to satisfy this condition by a small amount. Given the 
uncertainty in the definition of the field value at the new instanton core,
a minor modification of the profile can fix this. In any case, whether these two types of configurations are relevant or not for vacuum decay depends on their nucleation probability, which in turn depends on the potential considered, see  below.\\

{\bf $\bma{ \S\, 2}$ Approximate scale invariance.}
When there is no bounce, the false vacuum can still decay into an infinite number of bubble-like  configurations with $E=0$ but the integration over final states becomes crucial. Generically, the non existence of the Coleman bounce arises from perturbing a scale invariant potential ($V=-\lambda\varphi^4/4$ for QFT in 4 dimensions) for which there is a family of bounces with the same Euclidean action varying only in radius. 

This scale invariance can only be approximate: for the theory to have a ground state, the potential cannot be unbounded from below and one expects that some ultraviolet (UV) physics should create a potential minimum at some field value $\Lambda_{UV}$, which introduces a mass scale and breaks explicitly scale invariance. This UV physics might be unrelated to gravity, but gravity itself introduces an unavoidable mass scale, the Planck mass, and will also break scale invariance.  Moreover, radiative corrections cause $\lambda$ to be scale dependent and a more accurate evaluation of the potential should use $\lambda$ evaluated at a renormalization scale $\mu\sim\varphi$. This breaks scale invariance in a more subtle way and introduces a scale by dimensional transmutation. Finally, in some cases the quartic potential is just an approximation valid in some field regime and other operators (like a mass term) are present. 
The previous discussion is relevant for the Standard Model Higgs potential, which is approximately scale invariant with a negative quartic at high field values \cite{SMStab} and has all the sources of breaking mentioned.

The breaking of scale invariance by any of the sources described above lifts the degeneracy of the action for the family of Fubini bounces  (for $V\simeq -\lambda\varphi^4/4$) and three general outcomes can result: 1) The action is minimized by some field configuration (close to a Fubini form) with finite $\varphi(0)$ determined by the mass parameters breaking scale invariance (see e.g. \cite{ESM} for a recent discussion). This configuration dominates vacuum decay. 2) The action is minimized only asymptotically at $\varphi(0)\to\infty$ by bounces with radius tending to zero. In this case the rate is UV sensitive and requires knowledge of the UV completion of the model. 3) The action is minimized by $\varphi(0)\to 0$ (possible only if the potential has no barrier) by bounces with radius tending to infinity and is thus infrared dominated. Remembering that $\Gamma/V\sim R^{-4}e^{-S(R)}$, where $R$ is the bubble radius, we also see that 
the rate is suppressed at large bubble radius and the final rate
will come from a balance between prefactor and action, being maximized at the $R_\star$ satisfying $S'(R_\star)=-4/R_\star$.\footnote{Cases 2) and 3) can be illustrated by the potential
of (\ref{VmlL}). For $m^2\neq 0$ and $\Lambda\to\infty$ the action is minimized by instantons of zero radius and is thus UV dominated: introducing a finite $\Lambda$ gives a finite size bounce. On the other hand, for $m=0$ and finite $\Lambda$ the action is IR dominated, and introducing a finite (positive $m^2$) gives a finite size bounce.}

Breaking of scale invariance transforms the flat direction (for the action) in field configuration space into a sloping valley. This might
push the bounce (the minimum of the Euclidean action in the $E=0$ subspace) off to infinity (with either infinitely large or small bubble radius). In any case, if the slope of the valley is small, one needs to integrate over the valley to get the decay rate considering configurations that depart from the bounce. It is then that pseudo-bounces or new instantons might be relevant.\footnote{An alternative procedure used  to do this uses the so-called constrained instantons \cite{FY,Affleck}.}
\\

{\bf $\bma{ \S\, 3}$ Semiclassical approximation.}
The Euclidean bounce calculation of the false vacuum decay rate 
relies explicitly on the validity of the semiclassical approximation, which requires $S/\hbar\gg 1$. Being the decay a non-perturbative process, the action depends on some inverse coupling and the condition $S/\hbar\gg 1$ requires a weak coupling regime. In the semiclassical regime the rate is thus exponentially suppressed, reflecting the fact that the $E=0$ bubble configuration that needs to be created by field fluctuations over the false vacuum background is a large semiclassical object built of many quanta and thus expensive to form. In this sense, $S/\hbar$ measures how expensive the bubble is.
 
When the semiclassical approximation leads to $S/\hbar\simlt {\cal O}(1)$ we cannot calculate reliably the vacuum decay rate but we learn that the rate is fast (when the rate is slow, the semiclassical approximation does not fail). This is usually good enough as fast
decay rates just signal that the false vacuum in question cannot be of cosmological interest and a precise calculation of the decay rate is not needed.

The criterion of validity of the semiclassical approximation for the linear unbounded potential of section~\ref{sec:linear}  for large $\lambda_\mm$ and small $\lambda_\pp$ leads to the condition
\be
\lambda_\mm \ll \frac{32\pi^2}{3}\ ,
\label{lambdaweak}
\ee
obtained by requiring that the action (\ref{SC}) of the Coleman 
bounce is $S_C\gg 1$. Notice that (\ref{lambdaweak}) takes the form of a typical weak coupling condition on $\lambda_\mm$, even though perturbativity seems to allow for a large $\lambda_\mm$. In any case, for values of $\lambda_\mm$ that violate (\ref{lambdaweak}) one cannot trust any calculation that relies on the semiclassical approximation, either through bounces, pseudo-bounces or new instantons.

A similar condition (and conclusions) apply in the quartic potential of section~\ref{sec:quartic}. Imposing that the asymptotic value (\ref{Sasympt}) of the action satisfies the semiclassical condition gives
\be
\lambda_\mm \ll \frac{8\pi^2}{3}\ .
\ee

Instead of focusing on the bounce core, as done for new instantons in \cite{Mukhanov1,Mukhanov2,Mukhanov3,Mukhanov4}, we can apply to the whole bounce the criterium for classical  over quantum dominance requiring
\be
\Delta\varphi\gg\frac{1}{R}\ ,
\label{class}
\ee
where $\Delta\varphi\equiv |\varphi(0)-\varphi_0|$ is the jump in field values along the bounce and $R$ is the bounce radius. In the linear unbounded potential, with $\Delta\varphi\simeq 2\varphi_0$ and $R^2\simeq 8/(\lambda_\mm\varphi_0^2)$, condition (\ref{class})
translates into $\lambda_\mm\ll 32$, which is similar to the condition from large $S$ derived above. This is a generic result and makes physical sense: if the  bounce is quantum dominated, a fluctuation of typical amplitude can nucleate it and the decay rate is fast. Nucleating a semiclassical bounce requires many fluctuations going in the same direction and thus a large action.
\\

{\bf $\bma{ \S\, 4}$ Quantum fluctuations and rate prefactor.}
When the semiclassical approximation is valid, one can still worry about the impact of quantum fluctuations over the bounce \cite{CC}
which give a subleading correction to the rate and amount to a calculation of the non-exponential prefactor $A$ in $\Gamma/V=A e^{-S}$. A one-loop calculation of this prefactor, summing over Gaussian fluctuations over the bounce configuration, gives the well-known formula
\be
\frac{\Gamma}{V}=\left(\frac{S[\varphi_B]}{2\pi}\right)^2\left|\frac{\mathrm{Det} S''[\varphi_\pp]}{\mathrm{Det} ' S''[\varphi_B]}\right|^{1/2}\ e^{-S[\varphi_B]}\ ,
\label{prefactor}
\ee
where $S[\varphi_B]$ is the bounce action, $\mathrm{Det}$ refers to the functional determinant of the differential operators $S''$ which are the second variation of the action, evaluated on the false vacuum background $\varphi_\pp$ or the bounce $\varphi_B$ as indicated.
The prime in the last case indicates that (four translational) zero modes are removed. Integration over each zero mode gives a factor $\sqrt{S[\varphi_B]/(2\pi)}$,
which explains the $(S[\varphi_B]/(2\pi))^2$ coefficient in $A$.

 \begin{figure}[t!]
\begin{center}
\includegraphics[width=0.45\textwidth]{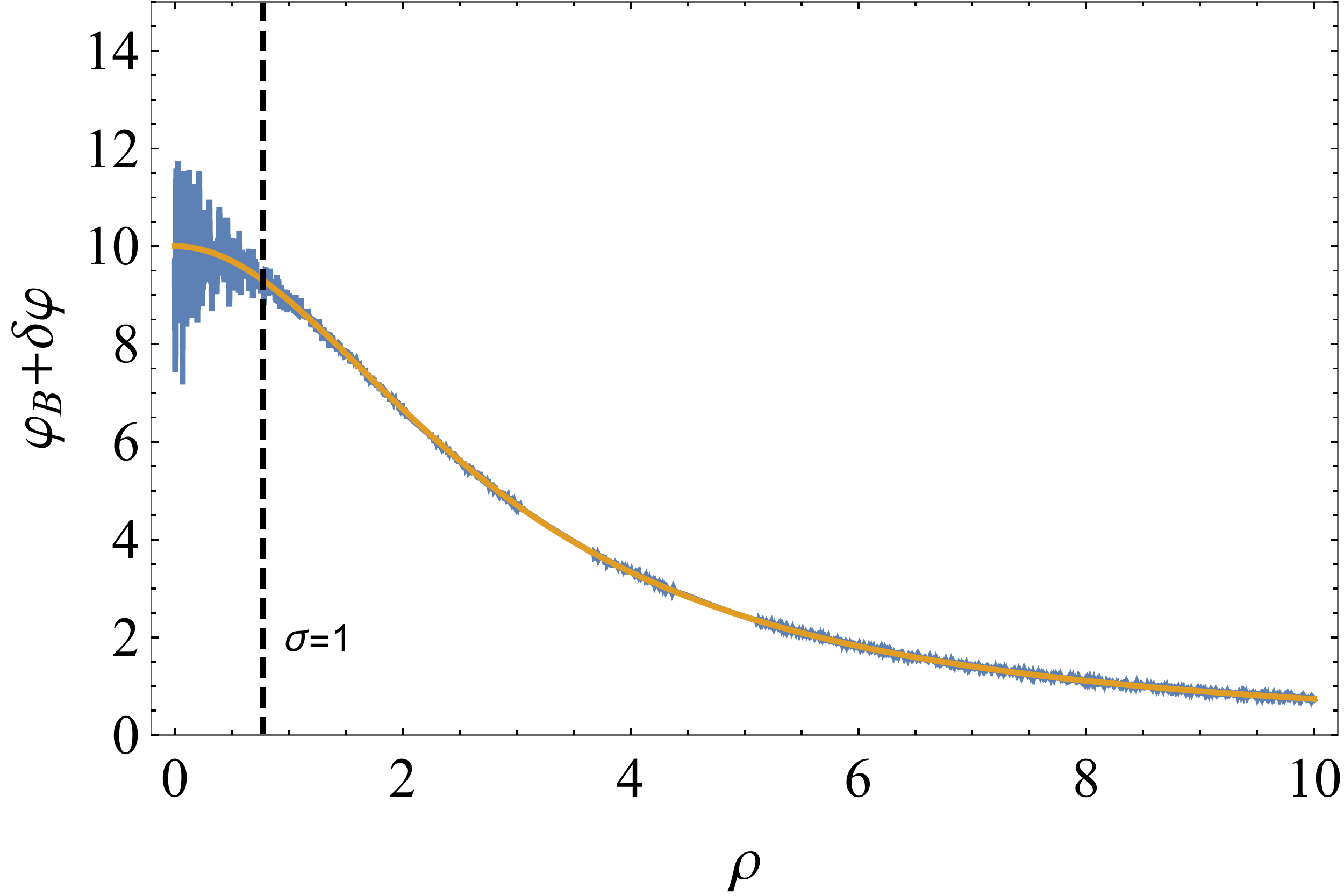}\hspace{2mm}
\includegraphics[width=0.45\textwidth]{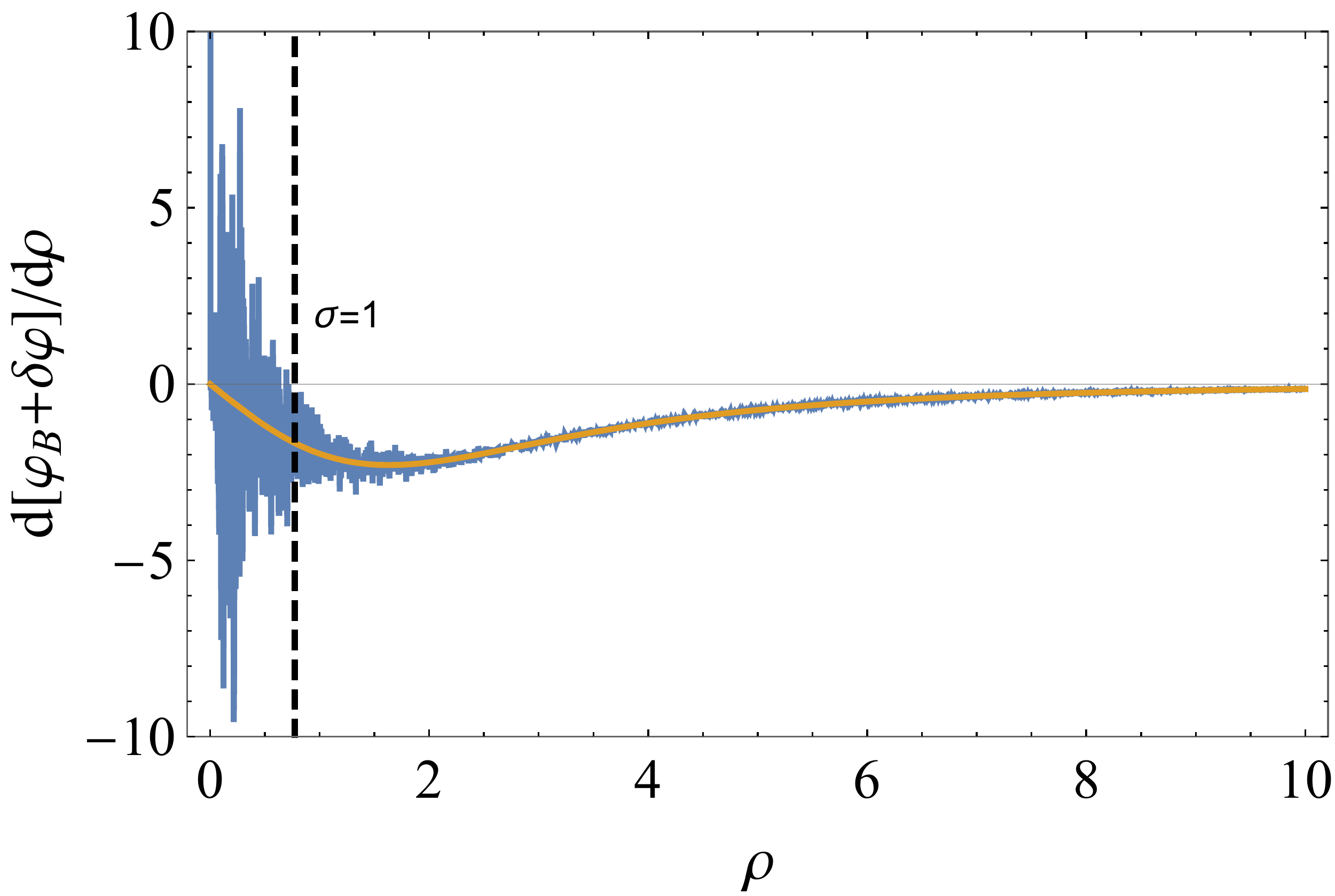}\\
\vspace{2mm}
\includegraphics[width=0.45\textwidth]{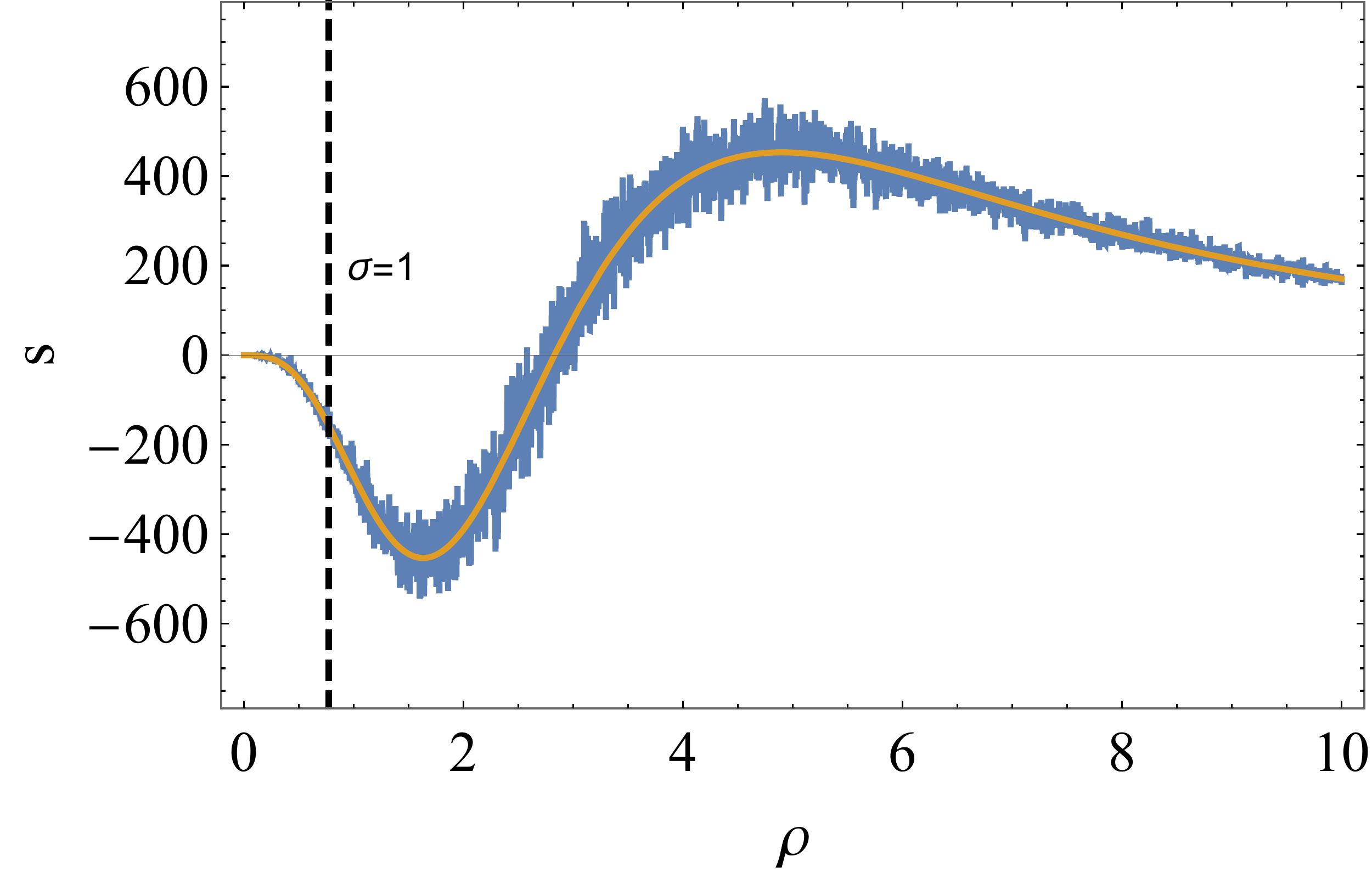}
\end{center}
\caption{Illustration of the impact of quantum fluctuations over the Fubini bounce (with $\varphi_e=10$ and $\lambda=0.01$) on the field (upper left plot), its radial derivative (upper right plot) and the action integrand (lower plot). In all cases the solid line is the semiclassical bounce without fluctuations. The black dashed line marks $\rho_{uv}$ for $\sigma=1$.
\label{fig:fluct}
}
\end{figure}

To get a feeling of the relative size of these quantum fluctuations over the bounce background, and for illustrative purposes, we can make use of the well studied case of the Fubini bounces, for which the full spectrum of the $S''[\varphi_B]$ operator is known. We take $V(\varphi)=-\lambda\varphi^4/4$ and follow the detailed analysis of \cite{Matt} (for similar discussions, see \cite{Prefactor0,Prefactor1}). We write the field corrected by quantum fluctuations as 
\be
\varphi(\rho) = \varphi_B(\rho) +\sum_j \xi_j \varphi_j(\rho)\ ,
\label{QField}
\ee 
where $\varphi_B(\rho)$ is the Fubini instanton,
$\varphi_B(\rho) =\varphi_e/(1+\rho^2/R^2)$,
with $1/R^2=\lambda\varphi_e^2/8$; $\varphi_j(\rho)$ are eigenfunctions of the (rescaled) operator ${\cal O}_\phi\equiv (-3\lambda\varphi_B^2)^{-1}\square-1$ with eigenvalues $\lambda_j$,
that we take to be normalized
as 
\be
\langle\varphi_j|\varphi_k\rangle\equiv -\int d^4x V''[\varphi_B]\varphi_j\varphi_k=2\pi \delta_{jk}\ ;
\ee
and $\xi_j$ are Gaussian variables with standard deviation $\sigma_j=1/\sqrt{2\pi\lambda_j}$. With this normalization, the path integral over each $\xi_j$ (with $\lambda_j\neq 0$) produces the
factor $1/\sqrt{\lambda_j}$ [leading to $\Pi_j \lambda_j^{-1/2}$ and the determinants in (\ref{prefactor})]. We illustrate these fluctuations in figure~\ref{fig:fluct} which shows the ``quantum corrected'' field (upper left plot), its derivative (upper right plot) and the action (lower plot) for $\varphi_e=10$ and $\lambda=0.01$. To generate these plots we used the eigenfunctions and eigenvalues given in \cite{Matt}, restricting the fluctuations to the radial ones, $\varphi_n$, with $n\leq 20$ and randomly sampling $\xi_n$ from a Gaussian distribution with $\sigma_n=1/\sqrt{2\pi\lambda_n}$. The dashed vertical line shows $\rho_{uv}$ for $\sigma=1$. Although such illustrations must be interpreted with care (for instance, one is ignoring non radial modes, the need of renormalization, etc.) they serve to show that a) the scale where quantum fluctuations become relatively large for the field and its derivative is correctly identified by $\rho_{uv}$; b) the importance of these fluctuations on the action is relatively small due to the $\rho^3$-suppression of the integrand in the core region, and c) the fluctuations at higher scales  are more important in the action.

In any case, to have a controlled calculation it is important to write the field as in (\ref{QField}) with a well defined semiclassical bounce over which to perturb. When the semiclassical expansion holds, the impact of quantum fluctuations is captured by the prefactor calculation and no special attention is needed to treat fluctuations near the center of the bounce. 
\\

{\bf $\bma{ \S\, 5}$ Pseudo-bounces or new instantons?}
In spite of the previous discussions, it is clear that new instantons (once $E=0$ is fixed) can also be the final state for vacuum decay
with an impact on the total rate that depends on their action. In sections~\ref{sec:linear} and \ref{sec:quartic} with particular examples and in section~\ref{sec:General} in general, we have seen that new instanton configurations are very close to those of pseudo-bounces but with actions that are larger, even if not by much. The difference
can be considered to be subleading, even smaller than the typical one coming from the prefactor calculation and this is due to the $\rho^3$-suppression of the action integrand, as seen in $\S\, 4$.
From this point of view, new instantons configurations would be accounted for by the prefactor calculation for pseudo-bounces.

 \begin{figure}[t!]
\begin{center}
\includegraphics[width=0.6\textwidth]{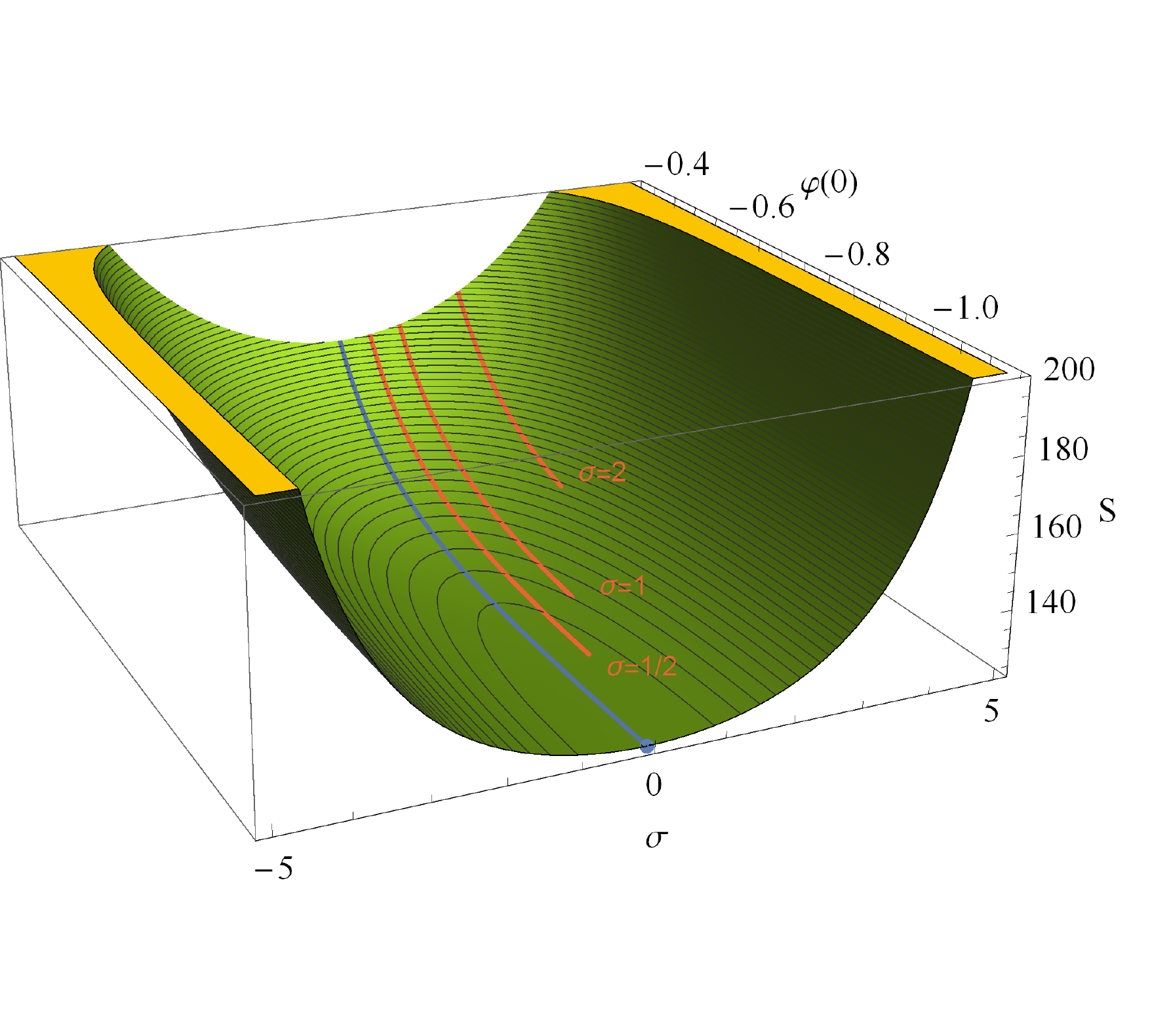}
\end{center}
\vspace{-1.2cm}
\caption{Tunneling action for the linear unbounded potential of section~\ref{sec:linear} with $\lambda_\mm=1$, $\lambda_\pp=0.1$ and $\varphi_0=1$, as a function of $\sigma$ and $\varphi(0)$. Pseudo-bounces are along the blue line in the valley bottom (ending at the Coleman bounce, marked by a dot) while new instantons are the red lines for the indicated values of $\sigma$ [ending at their respective lower limit on $\varphi(0)$].  (The $\sigma$ axis has been extended to unphysical negative values just to obtain a symmetric plot.)
\label{fig:SumPlot}
}
\end{figure}

The largest difference between the actions for these two configurations arises when new instantons are restricted to satisfy a limit on their core field value, that cannot be lower than a critical value. However, configurations that violate such limit (like pseudo-bounces do) are perfectly valid final states for vacuum decay and no physical principle nor selection rule forbids their nucleation.  Their importance for the rate is dictated by their action and whether quantum fluctuations have a large impact or not in such cases can be answered by the standard prefactor calculation. When the semiclassical approximation holds and the resulting action is large, one does not expect a large correction from quantum fluctuations, especially not from those coming from the bubble core. In such cases, the rate calculation via the new instantons with a lower limit on $\varphi(0)$ can underestimate the rate by orders of magnitude.

We conclude that, whenever the bounce fails to dominate the decay of a false vacuum, pseudo-bounces should be used rather than new instantons. The reasons are that pseudo-bounces track faithfully the bottom of the action valley in configuration space; they automatically have zero energy; and they are not restricted in field space so they can explore valid field configurations that can lower the decay
rate significantly. Figure~\ref{fig:SumPlot}, showing the action valley for the linear unbounded potential of section~\ref{sec:linear},  serves as a summary plot that captures the key points of the previous discussion.

\appendix
\section{WKB Decay in the Linear Unbounded Potential}

The Euclidean analysis of vacuum decay is ultimately justified by the  
WKB Minkowskian method in field theory, see \cite{BC}, which we follow below. In vacuum decay  the initial false vacuum configuration $\varphi_+=\varphi(\vec x,\alpha(t_1))$ (with $V_+=0$) tunnels to a zero-energy bubble configuration $\varphi(\vec x,\alpha(t_2))$ inside which the field probes the negative part of the potential. There is an energy barrier between these two configurations with a shape that depends on the path in configuration space, parametrized by $\alpha(t)$, that connects them. If we restrict the Minkowskian action for the scalar field
\be
S=\int d^3\vec x\, dt\left[\left(\frac{d\varphi}{dt}\right)^2-\frac12(\vec\nabla \varphi)^2-V(\varphi)\right]\ ,
\ee
to a tunneling path $\varphi_\alpha\equiv\varphi(\vec x,\alpha(t))$, we get
\be
S=\int_{t_1}^{t_2} dt \left[\frac12 m(\alpha)\left(\frac{d\alpha}{dt}\right)^2-{\cal V}(\alpha)\right]\ ,
\ee
where
\be
m(\alpha)\equiv\int d^3\vec x \left(\frac{d\varphi_\alpha}{d\alpha}\right)^2\ ,\quad
{\cal V}(\alpha)\equiv\int d^3\vec x \left[\frac12(\vec\nabla \varphi_\alpha)^2+V(\varphi_\alpha)\right]\ .
\ee
This reduces the problem to a one-dimension quantum mechanical tunneling problem and this works provided the semi-classical approximation holds. We can then use the  WKB expression
\be
S_{WKB}=2\int_{\alpha_1}^{\alpha_2}\sqrt{2m(\alpha){\cal V}(\alpha)}\ d\alpha\ ,
\ee
with $\alpha_i=\alpha(t_i)$ for the tunneling exponent for decay along this path. Vacuum decay will happen most likely via the path that minimizes this tunneling action.
As proven in \cite{BC}, $S_{WKB}$ above gives the Euclidean action result if Euclidean time is taken to satisfy
\be
\frac{d\tau}{d\alpha} = \sqrt{\frac{m(\alpha)}{2{\cal V}(\alpha)}}\ ,
\ee
and gets minimized for a path that is related to the Euclidean bounce solution $\varphi_B(\rho)$ by
\be
\varphi_\alpha = \varphi(\vec x,\alpha(t))=\varphi_B(\sqrt{\vec x^2+\tau^2})\ ,
\ee
making $O(4)$ invariance manifest.

Let us apply this formalism to the linear unbounded potential of section~\ref{sec:linear}. Consider a path in Minkowski space obtained from the Coleman bounce (\ref{CBL}) via the replacement
$\rho\rightarrow \sqrt{r^2+\alpha^2}$, where $r^2=\vec x^2$, 
\be
\varphi_\alpha(r,\alpha)\equiv 
\begin{cases}
\displaystyle{\frac18}\lambda_\mm\varphi_0^3(r^2+\alpha^2-\rho_{0,C}^2)\ ,
 & \mathrm{for}\quad r^2<\rho^2_{0,C}-\alpha^2\ ,\\
\varphi^{}_0\displaystyle{\frac{r^2+\alpha^2-\rho_{0,C}^2}{r^2+\alpha^2-\rho_{0,C}^2/(1+\delta_C)}}\ , & \mathrm{for}\quad r^2>\rho^2_{0,C}-\alpha^2\ ,
\end{cases}
\label{path2}
\ee
where $\rho_{0,C}^2=8(\lambda_\pp+\lambda_\mm)/(\lambda_\mm\varphi_0)^2$.
For $\alpha=0$ we get the $\tau=0$ slice of the bounce configuration and $\alpha=\infty$ gives the false vacuum $\varphi_0$. Notice that for $\alpha>\rho_{0,C}$ only the lower part in the definition (\ref{path2}) above is relevant.

It is then simple to get
\be
m(\alpha)=
\begin{cases}
%m_S(\alpha)\ ,
\displaystyle{
\frac{\pi\alpha^2\varphi_0^3}{\sqrt{2}\lambda_2^2}\left[\frac43\sqrt{\lambda_1}(2\lambda_1-\lambda_\mm)\left(\frac{\lambda_2^2}{\lambda_\mm}-\frac38\lambda_\mm+\lambda_2\right)+\frac{\lambda_\mm^3}{4}A(\alpha)\right]}\ ,
&
 \mathrm{for}\quad \alpha<\rho_{0,C}\ ,\\
 &\\
%m_L(\alpha)\ ,
\displaystyle{
\frac{\pi^2\alpha^2\lambda_\mm^3\varphi_0^3}{4\sqrt{2}(-\lambda_2)^{5/2}}}\ ,
&
 \mathrm{for}\quad \alpha>\rho_{0,C}\ .
\end{cases}
\ee
with
\be
\lambda_1\equiv \lambda_\mm+\lambda_\pp-\alpha^2\lambda_\mm^2\varphi_0^2/8\ ,\quad\quad
\lambda_2\equiv \lambda_\pp-\alpha^2\lambda_\mm^2\varphi_0^2/8\ ,
\ee
and
\be
A(\alpha)=
\begin{cases}
\frac{1}{\sqrt{\lambda_2}}\log\frac{1+\sqrt{\lambda_2/\lambda_1}}{1-\sqrt{\lambda_2/\lambda_1}}\ ,&
 \mathrm{for}\quad \lambda_2>0\ ,\\
\frac{1}{\sqrt{-\lambda_2}}\arctan\left(1+\lambda_2/\lambda_1,2\sqrt{-\lambda_2/\lambda_1}\right)\ ,
&
 \mathrm{for}\quad \lambda_2<0\ .\\
\end{cases}
\ee
Here we use the notation $\arctan(x,y)\in (-\pi,\pi)$ for the arc tangent of $y/x$ with the quadrant specified by $(x,y)$.

\begin{figure}[t!]
\begin{center}
\includegraphics[width=0.45\textwidth]{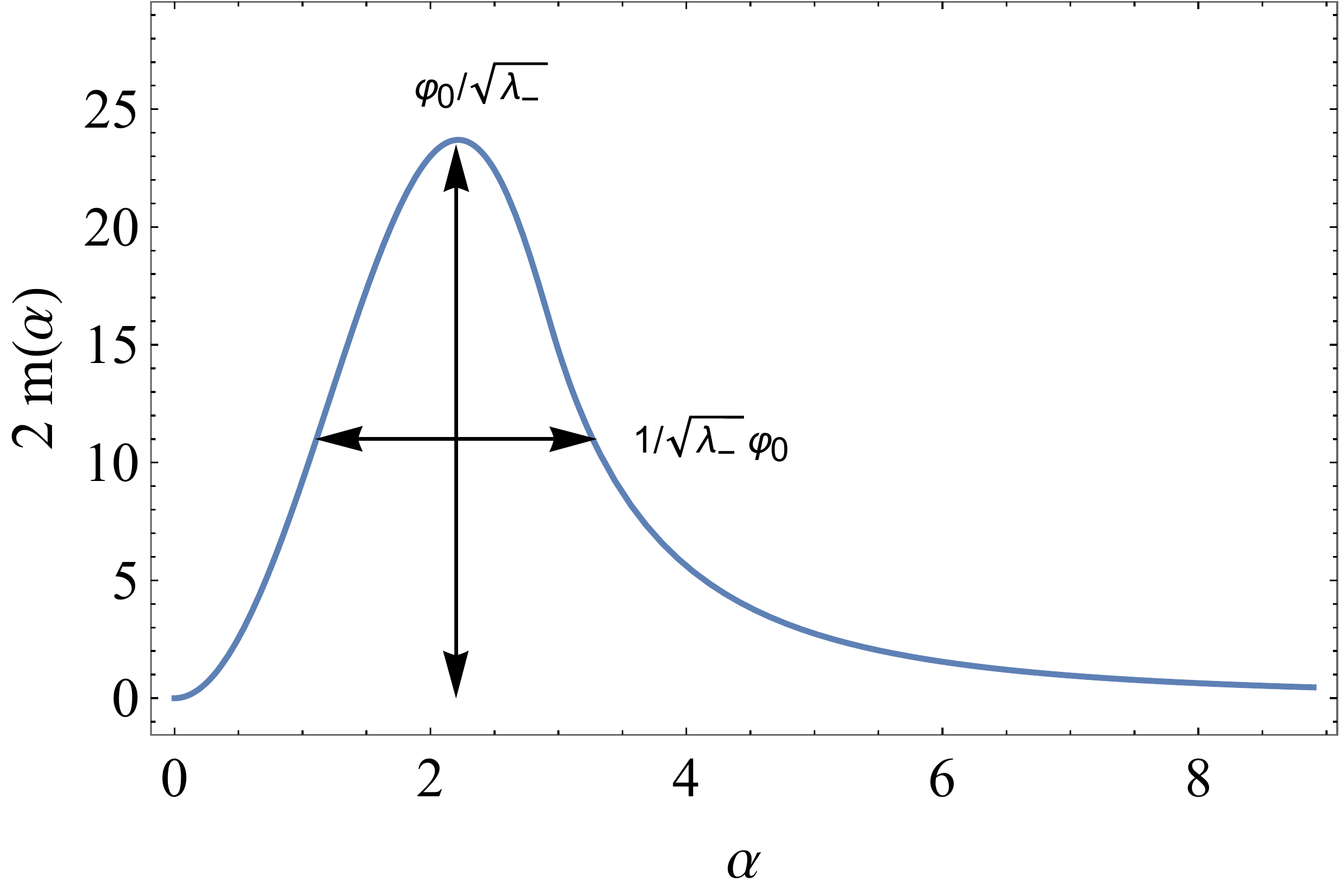}
\end{center}
\caption{Energy barrier for tunneling in the linear unbounded potential of section~\ref{sec:linear} for $\lambda_\pp=0.1$, $\lambda_\mm=1$ and $\varphi_0=1$. The parametric dependence 
of height and width is indicated. The area under the curve gives the tunneling action as in (\ref{SWKB}).
\label{fig:WKBarrier}
}
\end{figure}

With the simple $\alpha$ parametrizarion used, it can also be proven that  $m(\alpha)=2{\cal V}(\alpha)$.
Therefore
\be
S_{WKB}=\int_0^\infty 2 m(\alpha)d\alpha\ .
\label{SWKB}
\ee
Figure~\ref{fig:WKBarrier} shows the integrand above (for the same numerical example shown in figure~\ref{fig:actions}) as a function of $\alpha$. 
The integral for $S_{WKB}$ can be done analytically and the result reproduces the Euclidean one in (\ref{SC}).

In the limit of $\lambda_\mm\gg 1$ and small $\lambda_\pp$ one can get approximations for the height and width of the $2m(\alpha)$ energy barrier. Near the maximum of $2m(\alpha)$ we can use the expansion
\be
m(\alpha) = \frac{\pi\varphi_0}{\sqrt{\lambda_\mm}x^{3/2}}\left[\frac{(4-x)}{12}(x^2-8x-24)\sqrt{(8-x)x}+32\arctan\left(\frac{x-4}{x-8},\sqrt{\frac{x}{8-x}}\right)\right]+{\cal O}(\lambda_\pp)\ , 
\ee
with $x\equiv \lambda_\mm\alpha^2\varphi_0^2$.
It is then straightforward to get that $2m(\alpha)$ peaks at $x\simeq 4.6$ or
\be
\alpha_{max}\simeq \frac{2.1}{\sqrt{\lambda_\mm}\varphi_0}\ ,
\ee
and the peak has width of order $\alpha_{max}$ and
height
\be
2m(\alpha_{max})\simeq 40 \frac{\varphi_0}{\sqrt{\lambda_\mm}}\ ,
\ee
see figure~\ref{fig:WKBarrier}.
This shows that, for a fixed $\varphi_0$, the barrier gets lower and thinner with increasing $\lambda_\mm$. The resulting WKB action
scales roughly as [barrier width] $\times$ [barrier height] $\sim {\cal O}(10^2)/\lambda_\mm$, in agreement with the exact result.  

\section*{Acknowledgments.} 
This work has been supported by the Spanish Ministry for Science and Innovation under grant PID2019-110058GB-C22
and the grant SEV-2016-0597 of the Severo Ochoa excellence program of MINECO. 
%\bigskip

\end{document}